\documentclass[]{aa}
\usepackage{graphicx}
\usepackage{multirow}
\usepackage[varg]{txfonts}
\usepackage{xcolor}
\usepackage{booktabs}      
\newcommand{\kms}{km\,s$^{-1}$}

\newcommand{\msun}{M$_{\sun}$}
\usepackage{url}
\usepackage{relsize}

\begin{document}
\title{Submillimeter-wave emission of three Galactic red novae:\\ cool molecular outflows produced by stellar mergers}
\author{T. Kami\'nski\inst{\ref{inst1}}\thanks{Submillimeter Array Fellow}, 
        W. Steffen\inst{\ref{inst2}},
        R. Tylenda\inst{\ref{inst3}}, 
        K.H. Young\inst{\ref{inst1}}, N.A. Patel\inst{\ref{inst1}}, K.M. Menten\inst{\ref{inst4}}}
\institute{\centering 
    Harvard-Smithsonian Center for Astrophysics, 60 Garden Street, Cambridge, MA, USA, 
    \email{tkaminsk@cfa.harvard.edu} \label{inst1}
    \and Instituto de Astronom{\'i}a, OAN, UNAM, Ensenada, M{\'e}xico,\label{inst2}
    \and Nicolaus Copernicus Astronomical Center, Polish Academy of Sciences, Rabia{\'n}ska 8, 87-100 Toru\'n \label{inst3}
    \and Max-Planck-Institut f\"ur Radioastronomie, Auf dem H\"ugel 69, 53121 Bonn, Germany\label{inst4}}
\abstract{Red novae are optical transients erupting at luminosities typically higher than those of classical novae. Their outbursts are believed to be caused by stellar mergers. We present millimeter/submillimeter-wave observations with ALMA and SMA of the three best known Galactic red novae, V4332\,Sgr, V1309\,Sco, and V838\,Mon. The observations were taken 22, 8, and 14 yr after their respective eruptions and reveal the presence of molecular gas at excitation temperatures of 35--200\,K. The gas displays molecular emission in rotational transitions with very broad lines (full width $\sim$400\,\kms). We found emission of CO, SiO, SO, SO$_2$ (in all three red novae), H$_2$S (covered only in V838\,Mon) and AlO (present in V4332\,Sgr and V1309\,Sco). No anomalies were found in the isotopic composition of the molecular material and the chemical (molecular) compositions of the three red novae appear similar to those of oxygen-rich envelopes of classical evolved stars (RSGs, AGBs, post-AGBs). The minimum masses of the molecular material that most likely was dispersed in the red-nova eruptions are 0.05, 0.01, and 10$^{-4}$\,\msun\ for V838\,Mon, V4332\,Sgr, and V1309\,Sco, respectively. The molecular outflows in V4332\,Sgr and V1309\,Sco are spatially resolved and appear bipolar. The kinematic distances to V1309\,Sco and V4332\,Sgr are 2.1 and 4.2\,kpc, respectively. The kinetic energy stored in the ejecta of the two older red-nova remnants of V838\,Mon and V4332\,Sgr is of order $10^{46}$\,erg,  similar to values found for some post-AGB (pre-PN) objects whose bipolar ejecta were also formed in a short-duration eruption. Our observations strengthen the link between these post-AGB objects and red novae and support the hypothesis that some of the post-AGB objects were formed in a common-envelope ejection event or its most catastrophic outcome, a merger.}
\titlerunning{Red novae at submm wavelengths}
\authorrunning{T. Kami\'nski et al.}   
\maketitle

\section{Introduction}\label{intro}
The class of optical transients known as red novae are thought to be manifestations of stellar mergers happening in real time. As such, red novae and their remnants provide a glimpse into the extreme case of binary interaction taking place in common-envelope systems \citep{IvanovaCEE}. One distinguishing characteristics of red novae is that after the cataclysmic eruption the stellar remnant cools down to very low temperatures ($\sim$2000\,K) and its photosphere (or pseudo-photosphere) displays molecular bands typical for spectra of late-type supergiants. Their circumstellar medium also quickly cools down and copious amounts of molecular gas and dust are produced. Another distinct feature of red novae is that they erupt at luminosities of up to $\sim$10$^6$\,L$_{\odot}$, i.e. below those of supernovae but still brighter than classical novae. Red novae are thus considered a subgroup of intermediate luminosity optical transients (ILOTs) \citep{ILOTclass}. Owing to their relatively high luminosity, red novae can be observed in Local Group galaxies, with M31 LNR 2015 and M101 OT2015-1 being the two most recent examples of such transients \citep{kurtenkovM31,williamsM31,goranskijM101,M101,morgan}. Galactic red novae, whose remnants are in the focus of this study, include five objects: V4332\,Sgr, V838\,Mon, OGLE-2002-BLG-360, V1309\,Sco, and CK\,Vul.

From different vantage points, it becomes apparent that some objects classified as post-asymptotic giant branch (post-AGB) stars or pre-planetary nebulae (pre-PNe) could have been created in a red-nova like event (merger-burst) or, more generally, as an ILOT \citep[for classification, see][]{ILOTclass}. For instance, some transients classified as ILOTs, e.g. NGC300-OT 2008-1, could had been low-mass evolved stars prior to their eruption, just as expected for progenitors of post-AGB objects \citep{prieto}. \citet{jetsILOTs} link directly the formation of some planetary nebulae (PNe) with ILOTs and specifically propose that objects thus far classified as pre-PNe or post-AGB objects -- including OH231.8+4.2, IRAS 22036+5306, and M1-92 -- had formed in such events $\sim$10$^{2-3}$ years ago. Their conclusion is based on the fundamental physical characteristics of their outflows, such as kinematic age, mass, kinetic energy, momentum, and momentum rate.\footnote{Although a binary interaction is considered in the ILOT and PN  formation, the examples of Soker \& Kashi do not include \emph{mergers}.} On similar grounds, \citet{sahaiALMA} proposed a common envelope interaction and a subsequent merger of an evolved binary to explain unusual properties of the Boomerang Nebula, an object long considered to be a pre-PN with a bipolar outflow component. \citet{adiabatic} further suggested that the Boomerang Nebula was formed in an ILOT-like event. To add one more recent example, we note that based on the energy stored in the nebula of KjPn\,8, \citet{BM} suggested it formed in an ILOT event, or -- if it was indeed a binary merger -- in a red nova event. Many of these post-AGB and pre-PN objects are observed at millimeter and submillimeter (mm and submm) wavelengths in molecular lines which typically reveal bipolar outflows with remarkably high kinetic energy. Verifying whether any of these pre-PN erupted as an ILOT or a red nova in the past may prove to be very difficult or impossible. Inverting the problem, one may ask whether the known red novae will turn into objects that could be classified as post-AGB/pre-PN. In this paper, we present mm/submm observations of three Galactic red novae which allow first comparisons of their remnants to the pre-PNe and post-AGB objects. 

We investigate here remnants of the three Galactic red novae, V4332\,Sgr, V1309\,Sco, and V838\,Mon which are described in detail in Sect.\,\ref{sec-sources}. The merger remnants were observed using the Atacama Large (sub-)Millimeter Array (ALMA) and Submillimeter Array (SMA) with the aim of studying the physical and chemical characteristics of the cool circumstellar environment.  We describe and analyze the observations separately for each of the three objects in Sects.\,\ref{sec-V4332}--\ref{sec-V838}. The results are summarized and discussed in Sect.\,\ref{sec-discussion}.

\section{Galactic red novae}\label{sec-sources}
V838\,Mon erupted in 2002 and is perhaps the most recognizable red nova, also in popular media, owing to the spectacular light echo it displayed during and just after the eruption \citep{bond,tylendaEcho,echo}. The object belongs to a sparse open cluster \citep{cluster} partially embedded in a diffuse cloud of molecular gas and dust that remained after the cluster formation \citep{kami_coecho,echo}. This interstellar vicinity and the presence of lithium in the remnant are indicative of a young age of the V838\,Mon's progenitor, <25\,Myr. According to our best knowledge, prior to the 2002 eruption V838\,Mon belonged to a triple system dominated by two early B-type stars on the main sequence, each of about 8\,\msun\ \citep{tylendaProgenitor}, and separated by about 250\,AU \citep{tylFeII}. One of these B stars merged with a third member of the system, a proto-star of a mass of about 0.2\,\msun. The evolution of V838\,Mon during and just after the eruption is described in detail in \citet{tylEvolV838}. Polarimetric observations of the light echo surrounding V838\,Mon were used to derive a  distance of 6.1$\pm$0.6\,kpc and an outburst luminosity of about $6.5\times10^5$\,L$_{\sun}$ \citep{sparks} which makes it one of the most luminous red novae. Infrared (IR) interferometry suggests that the stellar remnant has been shrinking since the end of the outburst and in 2013 had a radius of 3.5$\pm$1.0\,AU. Spectra of the remnant show strong absorption lines of low-ionization atoms and broad absorption bands of metal-oxides. They indicate a cool oxygen-rich (O>C) circumstellar environment of the merger product, which had many characteristics of a red supergiant. The circumstellar medium observed at mid-IR exhibits an eccentric structure with a maximum size of 150--400\,AU \citep[depending on the wavelength;][]{olivier}. It likely has a molecular component, seen so far mainly in bands of H$_2$O. It is unclear, whether it is a disk-like structure seen at a high inclination or a bipolar outflow. In 2005, part of the material lost in the eruption reached the B-type companion of V838\,Mon and completely obscured the hot star  \citep{munariEclipse,tylFeII}. The cool stellar remnant of V838\,Mon, on the other hand, has been increasing its optical brightness in the last decade (V. Goranskij\footnote{\url{http://www.vgoranskij.net/v838mon.ne3}}), possibly owing to the continuing contraction. Soon after the eruption, an SiO maser was found toward V838\,Mon \citep{deguchi,claussen,DeguchiATel} and can be still observed today. Together with other spectral features in the visual spectrum \citep{kamiKeck}, the SiO maser indicates ongoing mass loss, e.g. a wind, from the late-M type supergiant. Simultaneously, some material may be falling back on the stellar remnant \citep{rushton,geballe,tylFeII}. V838\,Mon is the best observed red nova and its appearance triggered new interest in optical transients and stellar-merger eruptions. The foundations to interpret V838-Mon-like eruptions as stellar mergers were laid in \citet{TS06}.

V1309\,Sco erupted as a red nova in 2008 \citep{mason}. The eruption at maximum light was one order of magnitude less luminous than that of V838\,Mon \citep{v1309}. While pre-outburst observations for most red novae are very sparse, V1309\,Sco was frequently observed by OGLE \citep{ogle} since 2001, i.e. over about six years preceding the eruption. The OGLE light curve indicates that V1309\,Sco was an eclipsing contact binary system with an orbital period of $\sim$1.4 days. Moreover, it is evident in the OGLE data that the orbital period was slowly decreasing, providing, for the first time, a direct view on spiraling in of two stars into a merger. The stars that coalesced were first-ascension giants of early spectral types K and a total mass of $\sim$1.7\,\msun; their  initial orbital period was 2.5--3.1 days and the zero-age main sequence masses were of 1.3+0.5\,\msun\citep{v1309,stepien}. There is also evidence that this evolved binary system was losing mass just before the merger, mainly in the orbital plane \citep{v1309,McCollum,v1309sed,PejchaBipolar}. The stellar product of the merger has not been so far observed directly but is now at least 50 times fainter than in the maximum light \citep{v1309sed}. Its circumstellar remnant is rich in cool molecular gas and dust formed before and during the eruption \citep{nicholls,v1309sed}. The molecular component is oxygen-rich and is manifested by strong emission features of metal oxides, including AlO and CrO which are very rarely observed in astronomical objects  \citep{kamiV1309}. Currently, V1309\,Sco is a very challenging source for visual observations owing to a very low brightness and field crowding. It is however a relatively bright infrared source \citep{v1309sed}.

V4332\,Sgr was discovered in outburst in 1994 \citep{martini}. Its eruption was typical for red novae and produced a red supergiant embedded in cool circumstellar environment. Although the remnant has not been spatially resolved, spectroscopic and spectropolarimetric observations complemented by an SED analysis indicate that the supergiant is surrounded by a disk-like structure seen nearly edge on and associated with a bipolar outflow which may be (i) a wind from the central star, (ii) a disk wind, or (iii) material lost during the eruption \citep{kamiV4332_2010,KamiPol,KamiSpecpol,tyl2015}. Dust was formed abundantly also in this red nova and some grains have icy mantles, perhaps providing an evidence for a long history of mass loss from before the outburst \citep{baner04}. Some material may be falling back on the star which itself became $\sim$300\,K cooler in 2007 and is now an M5-6 supergiant. The remnant has a unique visual and infrared spectrum dominated by strong emission lines of neutral alkali metals and bands of simple oxides, including AlO and CrO \citep{kamiV4332_2010,tyl2015}. Remarkably, the only other source with a similar SED and spectra is the remnant of the younger red nova V1309\,Sco \citep{kamiV1309}. Not much is known about the progenitor of V4332\,Sgr but the striking similarity of its remnant to that of V1309\,Sco may indicate it could have been a similar system as that seen by OGLE in V1309\,Sco. 

Two other objects, CK\,Vul and OGLE-2002-BLG-360, have been proposed to belong to the red novae class \citep{kato,tyl-blg360} but are not included in the current study. The cool environment of CK\,Vul has been extensively studied at submm/mm wavelengths and is described in separate papers \citep{kamiNat,kami_ckvuldish}. BLG-360, on the other hand, has not yet been characterized through spectroscopic observations. In particular, submm observations of this object have not been attempted. We come back to these two red novae at the end of this paper (Sect.\,\ref{sec-discussion}).

In the following sections, we describe the new mm and submm observations of the three best-studied red novae.
\section{V4332 Sgr}\label{sec-V4332}
\subsection{ALMA observations}\label{sec-obs-v4332}
V4332\,Sgr was observed with ALMA on 12 and 13 August 2016 and on 3 July 2017 with 37, 40, and 44 antennas, respectively. The baselines ranged from 15.1\,m to 2.4\,km providing us with a beam FWHM of 168$\times$122\,mas with Briggs weighting and the robust parameter $R$=0.5; the largest recoverable scale of 7\farcs0 is much larger than any emission region observed in V4332\,Sgr and hence the data do not suffer from the missing-flux problem. The bandpass and flux calibration was performed by observations of J1924-2914. The complex-gain calibrators were quasars J1838-1853 and J1911-2006 in 2016, and J1832-2039 and J1845-2200 in 2017. The data were calibrated with the default ALMA CASA pipeline version 4.7.2. 

The observations were arranged in four Band\,7 spectral windows, each with a bandwidth of 1.875\,GHz and a channel width of 7812.5\,kHz. The spectra covered 344.2--347.9 and 356.1--359.9\,GHz with a small gap in 357.95--358.04\,GHz. The spectral setup is almost the same as that for V1309\,Sco (Sect.\,\ref{sec-obs-v1309}). 

No continuum was detected at an rms of 35\,$\mu$Jy/beam. In the data for which the gains were calibrated on the quasars, the cumulative emission within the entire band resulted in a S/N of 157 (maximum pixel value divided by the map rms with Briggs weighing at $R$=0.5). The source of the cumulative molecular emission is located at $\alpha$=18$^h$50$^m$36\fs69258 ($\pm$1.5\,mas) and $\delta$=--21\degr23\arcmin29\farcs0237 ($\pm$1.2\,mas)\footnote{The uncertainties quoted for position measurements in this paper express only the statistical error related to the fit of an elliptical Gaussian to the observed emission region. Astrometric uncertainties related to the calibration of phases typically add a systematic error of over 3\,mas for the ALMA data presented here (ALMA Technical Handbook).} (J2000). This value is at a small offset from the original phase center set to the position of V4332\,Sgr in the 2MASS catalog  but still consistent within the uncertainties with all earlier position determinations for V4332\,Sgr, including that of \citet{martini} ($\alpha$=18$^h$50$^m$36\fs73 and $\delta$=--21\degr23\arcmin28\farcs98, $\pm$0.26\,arcsec). Further improvement in the calibration of phases was performed using the emission peak of the strongest observed line, SiO $J$=8--7, as the reference model in CLEAN. This increased the S/N to 200 at the cost of the absolute astrometric information.
 		 	 	
The ALMA observations were executed 1 day and 11 months apart. The line fluxes determined from data taken in the two consecutive days are consistent within 2.2\%. Although, the observations at the two extreme epochs were made with two different array configurations, both resulted in very satisfactory coverages of the {\it uv} plane and their corresponding source fluxes can be compared directly. Spectra extracted for the entire source show that the flux had dropped nearly equally in all molecular features, on average by 18\%, over the 11 months. Although this value is larger than the formal 3$\sigma$ uncertainty in the ALMA absolute flux calibration (15\%), the systematic character of this decrease indicates it is not physical and is caused by uncertainties in the calibration. We therefore assume here that the actual flux calibration is at a level of 18\%.  The three datasets were combined. 

\begin{figure*}
\includegraphics[angle=270,width=\textwidth]{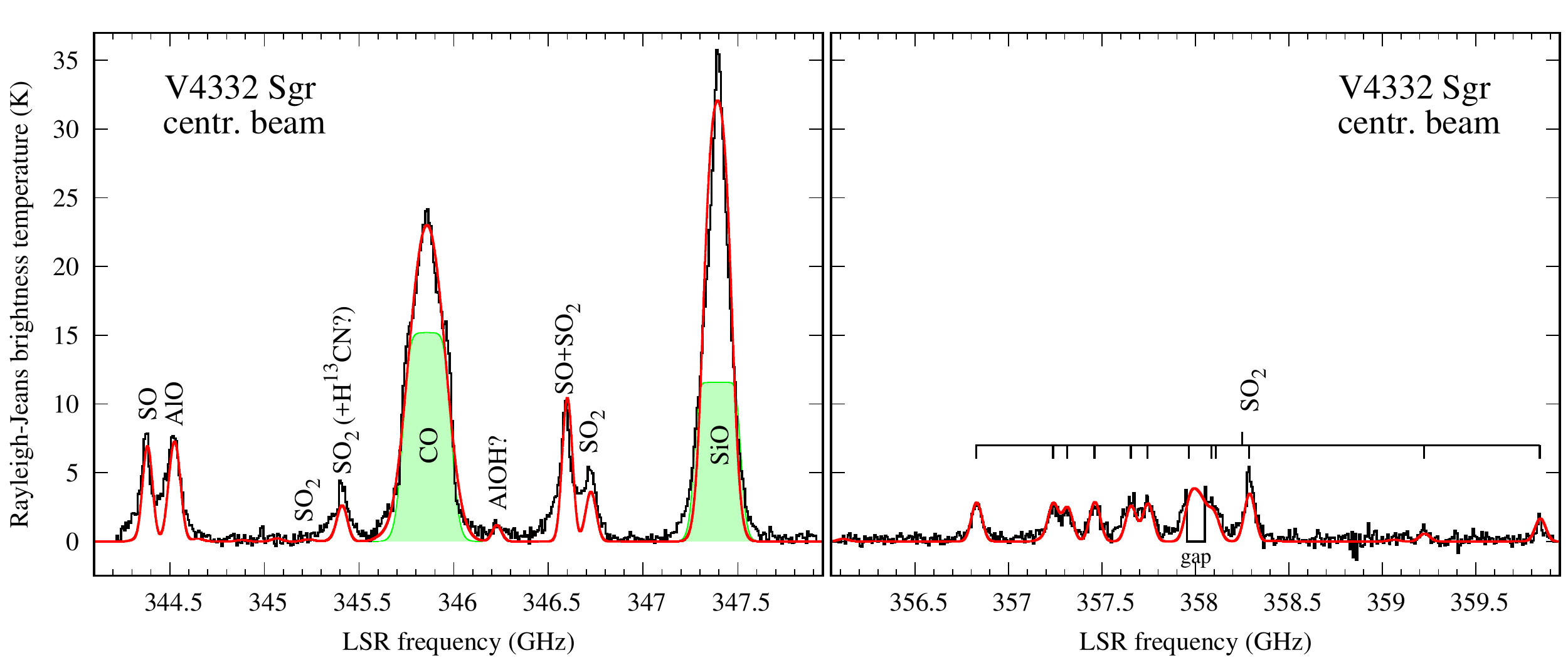}
\caption{ALMA spectrum of V4332\,Sgr and its simulations. The observed spectrum is shown with black lines and was extracted for the central beam of the molecular region. The spectrum is expressed in the Rayleigh-Jeans brightness temperature. Our best CASSIS simulation of all the spectral features at a single excitation temperature of 93\,K is shown with a red line. The green line and shaded areas show model SiO and CO emission excited in gas of a temperature of 34\,K; the profiles are very saturated and the column densities cannot be reliably constrained. All spectral features seen in the right panel are of SO$_2$.}\label{fig-cassisModel-V4332}
\end{figure*}

\subsection{Identification of spectral features}\label{sec-ident_v4332}
The identification of spectral features was performed in CASSIS\footnote{\url{http://cassis.irap.omp.eu}} and was based on standard catalogs of mm/submm lines, mainly on the Jet Propulsion Laboratory catalog\footnote{\url{https://spec.jpl.nasa.gov}} \citep{jpl}. The identified lines are listed in Table\,\ref{tab-lines} (see also Fig.\,\ref{fig-cassisModel-V4332}).

\begin{table*}\caption{List of lines detected in red novae.}\label{tab-lines}\centering\small
\begin{tabular}{cccr ccc}
\hline
Molecule & Quantum & Rest frequency & \multicolumn{1}{c}{$E_u$} &  \multicolumn{3}{c}{Total line flux (Jy \kms)} \\ %
         & numbers & (MHz)          & \multicolumn{1}{c}{(K)}   & V4332\,Sgr   & V1309\,Sco   & V838\,Mon\\ %
\hline
$^{29}$SiO&(5--4)          & 214385.7520 &  30.87 & \tablefootmark{c} & \tablefootmark{c} & 14.4 \\
SO$_2$  &(16,3,13--16,2,14)& 214689.3941 & 147.84 & \tablefootmark{c} & \tablefootmark{c} &  2.6 \\
SO      &(5,5--4,4)        & 215220.6530 &  44.10 & \tablefootmark{c} & \tablefootmark{c} &  6.5 \\
SO$_2$  &(22,2,20--22,1,21)& 216643.3035 & 248.45 & \tablefootmark{c} & \tablefootmark{c} &<4.9\tablefootmark{a} \\
H$_2$S  &(2,2,0--2,1,1)    & 216710.4365 &  83.98 & \tablefootmark{c} & \tablefootmark{c} &<4.9\tablefootmark{a} \\
SiO     &(5--4)            & 217104.9190 &  31.26 & \tablefootmark{c} & \tablefootmark{c} & 20.6 \\
CO      & (2--1)           & 230538.0000 &  16.60 & \tablefootmark{c} & 3.3               & 21.8 \\ %
$^{29}$SiO& (7--6)         & 300120.4801 &  57.62 & \tablefootmark{c} & \tablefootmark{c} & 20.9 \\
H$_2$S  &(3,3,0--3,2,1)    & 300505.5600 & 168.90 & \tablefootmark{c} & \tablefootmark{c} &  5.6 \\
SO      &(7,7--6,6)        & 301286.1240 &  70.96 & \tablefootmark{c} & \tablefootmark{c} &  5.5 \\
SO$_2$  &(19,2,18--19,1,19)& 301896.6287 & 182.63 & \tablefootmark{c} & \tablefootmark{c} &  2.5 \\
$^{30}$SiO&(8--7)          & 338930.0580 &  73.20 & \tablefootmark{c} & \tablefootmark{c} & 37.5 \\
\multicolumn{6}{l}{gap}\\ %
SO  & (8,8--7,7)           & 344310.6120 &  87.48 & 2.0  & <4.8 & \tablefootmark{c}\\ %
AlO & (9--8)               & 344451.9400 &  82.77 & 3.0  & <4.8 & \tablefootmark{c}\\ %
SO$_2$& (5,5,1--6,4,2)     & 345148.9708 &  75.15 & <0.3 & $\lesssim$0.7 & \\ %
SO$_2$& (13,2,12--12,1,11) & 345338.5377 &  92.99 & 1.5  & $\lesssim$1.3 & \\ %
CO  & (3--2)               & 345795.9899 &  33.19 & 27.6 & 9.6 & 42.1 \\ %
AlOH&(11--10)              & 346160.8387 &  99.70 & <0.6?&     &       \\ 
SO$_2$& (16,4,12--16,3,13) & 346523.8784 & 164.47 &$\lesssim$3.2\tablefootmark{a} &<0.6\tablefootmark{a} \\
SO  & (8,9--7,8)           & 346528.4810 &  78.78 &$\lesssim$3.2\tablefootmark{a} &<0.6\tablefootmark{a} & $\lesssim$19.7\tablefootmark{a} \\ 
SO$_2$& (19,1,19--18,0,18) & 346652.1691 & 168.14 &$\lesssim$1.8 & <6.0 & $\lesssim$19.7\tablefootmark{a}\\ 
SiO & (8--7)               & 347330.5810 &  75.02 & 22.2 & 6.7 & 47.3 \\ 
\multicolumn{6}{l}{gap}\\ %
SO$_2$& (10,4,6--10,3,7)   & 356755.1899 &  89.84 & 0.9           & <0.9\tablefootmark{a} & \tablefootmark{c}\\ 
SO$_2$& (13,4,10--13,3,11) & 357165.3904 & 122.97 & $\lesssim$1.0\tablefootmark{a} & <2.2\tablefootmark{a} & \tablefootmark{c}\\ 
SO$_2$& (15,4,12--15,3,13) & 357241.1932 & 149.68 & $\lesssim$1.0\tablefootmark{a} & <2.2\tablefootmark{a} & \tablefootmark{c}\\ 
SO$_2$& (11,4,8--11,3,9)   & 357387.5795 &  99.95 & 0.9           & <3.7\tablefootmark{a} & \tablefootmark{c}\\ 
SO$_2$& (8,4,4--8,3,5)     & 357581.4486 &  72.36 & $\lesssim$1.6\tablefootmark{a} & <3.7\tablefootmark{a} & \tablefootmark{c}\\ 
SO$_2$& (9,4,6--9,3,7)     & 357671.8206 &  80.64 & $\lesssim$1.6\tablefootmark{a} & <4.7\tablefootmark{a} & \tablefootmark{c}\\ 
SO$_2$& (7,4,4--7,3,5)     & 357892.4422 &  65.01 & $\gtrsim$0.7\tablefootmark{b} & <4.7\tablefootmark{ab} & \tablefootmark{c}\\
\multicolumn{6}{l}{gap}\\ %
SO$_2$& (5,4,2--5,3,3)     & 358013.1536 &  53.07 & <1.1\tablefootmark{ab} & <4.2\tablefootmark{ab} & \tablefootmark{c}\\
SO$_2$& (4,4,0--4,3,1)     & 358037.8869 &  48.48 &<1.1\tablefootmark{a} & <4.2\tablefootmark{a} & \tablefootmark{c}\\ 
SO$_2$& (20,0,20--19,1,19) & 358215.6327 & 185.33 & 1.1 & <4.2\tablefootmark{a} & \tablefootmark{c}\\ 
SO$_2$& (25,3,23--25,2,24) & 359151.1581 & 320.93 & 0.2 &  0.6 & \tablefootmark{c}\\ 
SO$_2$& (19,4,16--19,3,17) & 359770.6846 & 214.26 & 0.2 &  1.0 & \tablefootmark{c}\\ 
\hline
\end{tabular}\tablefoot{
Total line fluxes represent the entire emission region. Only very approximate fluxes are given for weak and blending features.
\tablefootmark{a} Tight blend. 
\tablefootmark{b} Line partially trimmed by the gap.
\tablefootmark{c} Line was not covered in observations.
}
\end{table*}

We first identified the two anticipated  lines, CO $J$=3--2 and SiO $J$=8--7, which are commonly present in cool circumstellar envelopes. This allowed us to constrain the source central velocity and identify weaker lines. From visual observations it was established that the source is oxygen rich \citep{tylV4332_2005,kimeswenger,kamiV4332_2010}, i.e. the molecular species expected in this source are characteristic of chemistry at a C-to-O ratio $<1.0$. We thus compared the spectrum of V4332\,Sgr to these of classical O-rich circumstellar envelopes \citep[e.g.][]{kami_surv} and found 15 lines of SO$_2$, 2 lines of SO, and a single feature of AlO and AlOH. Because of a very close match in velocity and previous identifications in visual spectra of V4332\,Sgr \citep{kamiV4332_2010,tyl2015}, we are confident that these identifications are reliable even if only one line of the species is covered. One exception is AlOH whose single feature ascribed to $N$=11--10 is observed at a very modest S/N and the identification is only tentative. The numerous lines of SO$_2$ are relatively broad (e.g. compared to analogues lines in AGB stars) and often located close to each other so that they often blend. The covered transition of AlO $N$=9--8, has a considerable hyperfine splitting that increases the apparent width of the observed feature near 344.45\,GHz \citep[cf.][]{kamiAlOmira}. There are no strong unidentified lines but some of the identified ones appear to be contaminated by other unassigned weak emission features (or have irregular line profiles). For instance, based on excitation analysis of the SO$_2$ spectrum (Sect.\,\ref{sec-cassis-v4332}), the observed line $J,K_a,K_c$=13,2,12--12,1,11 near 345.3\,GHz is stronger than expected in local thermal equilibrium and may be contaminated by emission of H$^{13}$CN $J$=4--3 that has its rest frequency within 1.1\,\kms. Without observations of other transitions of H$^{13}$CN or HCN, however, we are not confident about the presence of the 4--3 line. 

\subsection{Location and size}\label{sec-v4332-sizes}
The central position of the emission of SiO, CO, and the SO+SO$_2$ blend near 346.65\,GHz were investigated by fitting a Gaussian to their total-intensity maps (before self-calibration). The locations are consistent, within uncertainties, with each other and with the average position determined for integrated flux of all emission features (Sect.\,\ref{sec-obs-v4332}). The sizes of the regions corresponding to the different species are, however, different. 

\begin{figure*}[!h]
\sidecaption
\includegraphics[width=6cm,trim={10 10 50 10},clip]{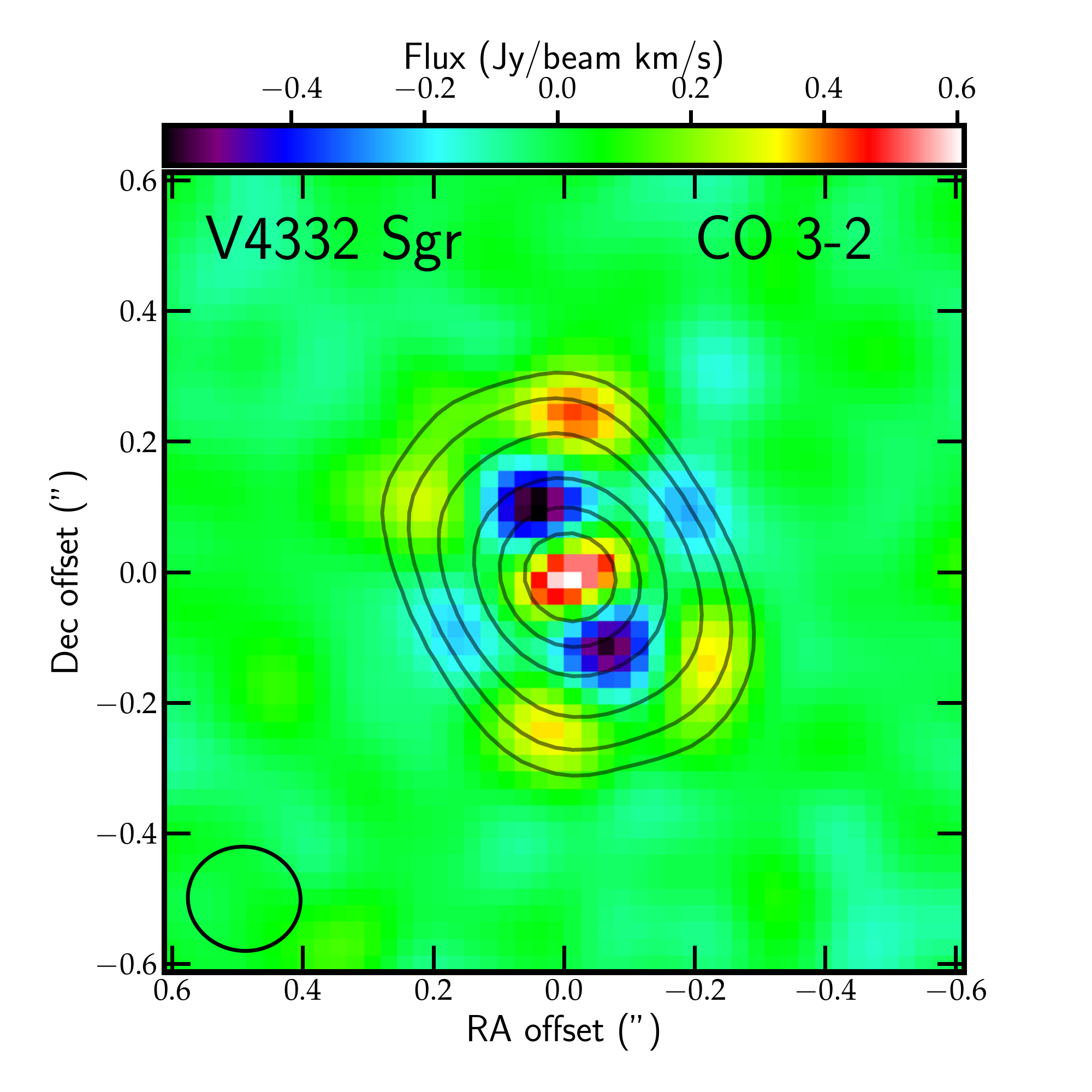}
\includegraphics[width=6cm,trim={10 10 50 10},clip]{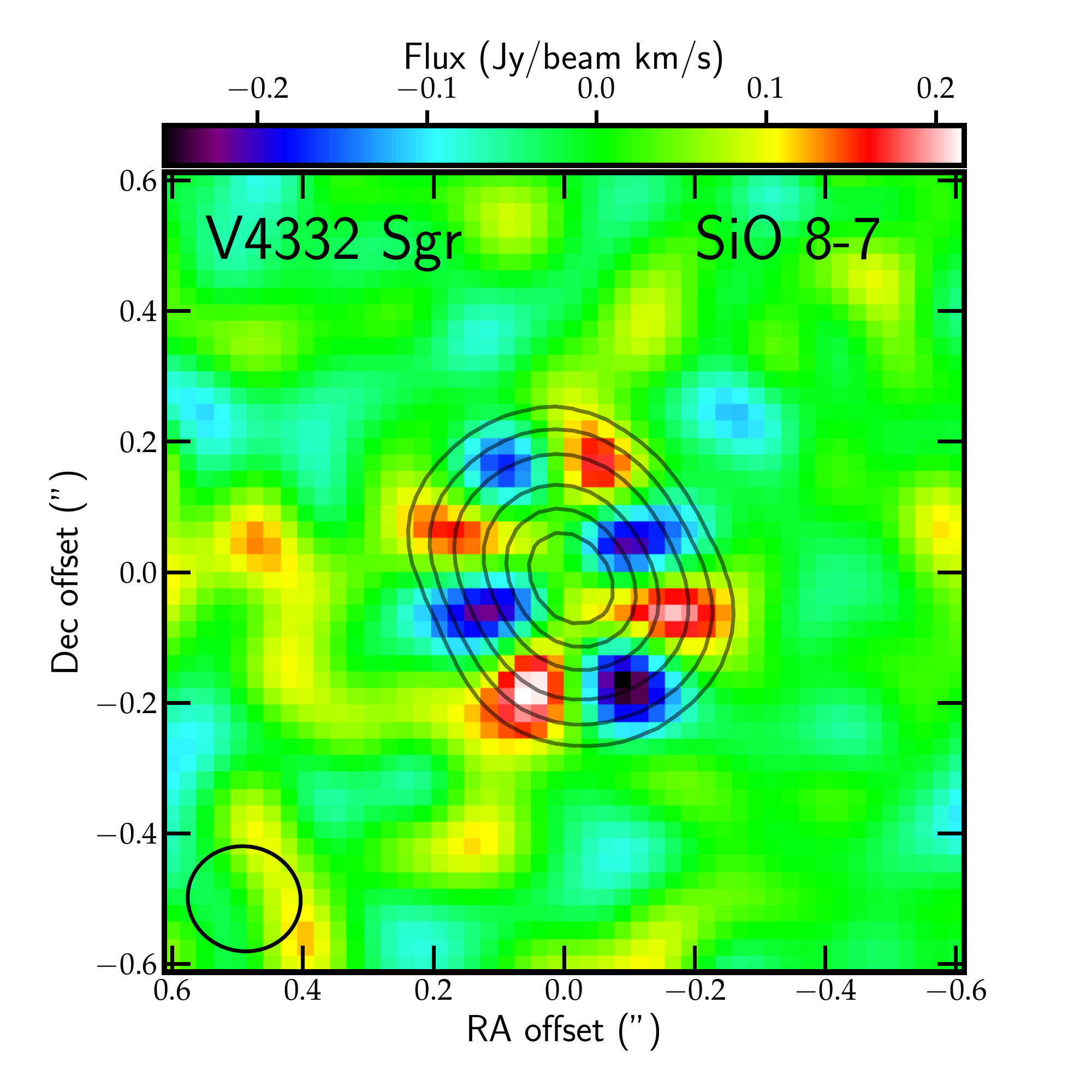}
\caption{Maps of molecular emission in V4332\,Sgr. The contours show the total (velocity-integrated) line flux of CO 3--2 (left) and SiO 8--7 (right) and the background image shows residuals after subtraction of a single best-fit elliptical Gaussian from the total intensity maps. The contours are drawn at 5, 10, 20, 40, 60, and 80\% of the maximum flux (12.10 and 12.06 Jy/beam\,\kms\ for CO and SiO, respectively). The ALMA beam sizes are shown in lower left corners.}\label{fig-residuals}
\end{figure*}

We measured the beam-deconvolved sizes by finding a Gaussian profile best representing the emission region. The emission of CO and SiO is indeed most extended and has bipolar or even quadruple substructure that cannot be represented by a single Gaussian component. The substructure is clearly apparent in the maps of residuals shown in Fig.\,\ref{fig-residuals}. Nevertheless, the single-component fits yield rough constraints on the sizes of the emission regions of 232$\times$165 ($\pm$6) mas for CO and of 189$\times$116 ($\pm$3) mas for SiO (all sizes were corrected for the beam smearing). Other emission regions are smaller, e.g. we get 157$\times$72 ($\pm$6) mas for AlO and 114$\times$91 ($\pm$5) mas for the blend of SO and SO$_2$ near 346.65\,GHz. The longer axis of these emission regions are all consistently aligned at a position angle of 31\fdg1$\pm$1\fdg4 (weighted mean and 1$\sigma$ uncertainty). The only exception is the emission of SO$_2$ integrated in multiple transitions in the range 357.00--358.42\,GHz: its emission region is circular and has a deconvolved FWHM of 109$\pm$10\,mas. 

The overall orientation of the longer axis of the emission regions at 31\fdg1$\pm$1\fdg4 is perpendicular to the polarization vector measured in the visual continuum of 114\fdg6$\pm$2\fdg0. This relative alignment can be easily understood within the model proposed in \citet{KamiPol} and \citet{KamiSpecpol}. Based on visual and infrared observations, which did not resolve the source, they found that the cool circumstellar gas of V4332\,Sgr has a morphology of a bipolar wide-angle outflow expanding perpendicular to a dusty disk seen almost edge-on and obscuring the stellar remnant. The observed high polarization degree of 17\% of the visual continuum is produced through efficient scattering that takes place in the bipolar outflow above and below the plane of the disk, i.e. at scattering angles of $\sim$90\degr. The difference in the position angles of the polarization vector and orientation of the submm emission of 83\fdg5$\pm$3\fdg5 is within 2$\sigma$ uncertainties equal to 90\degr, as expected if the emission regions mapped by ALMA trace the same cool gas as seen in the visual and infrared observations. The ALMA observations provide the first opportunity to study directly the spatio-kinematic structure of the bipolar remnant in spatially-resolved maps. 

\subsection{Overall kinematics and distance}\label{sec-profiles-distance}

\begin{figure}[!h]
\includegraphics[width=0.45\textwidth]{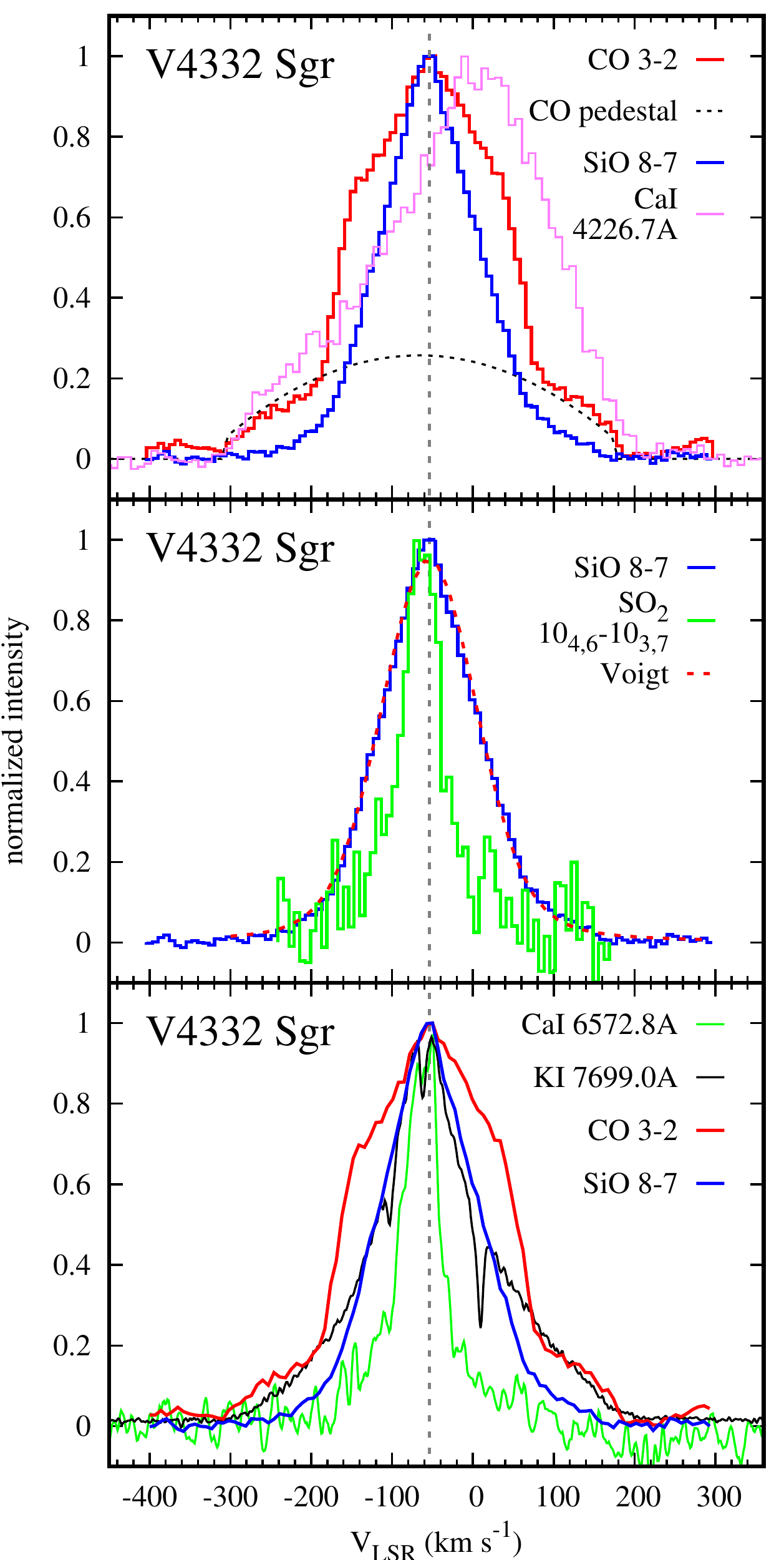}
\caption{Sample emission profiles in V4332\,Sgr representing the entire emission region. {\bf Top:} The profiles of rotational emission of SiO 8--7 and CO 3--2. The black dashed line shows the best-fit parabole representing the high-velocity component. {\bf Middle:} The profile of SiO is compared to a Voigt profile and the narrower line of SO$_2$ $10_{4,6}$--$10_{3,7}$ is also shown. {\bf Bottom:} The two ALMA lines from the top panel are compared to electronic transitions of \ion{K}{I} (at 7698.974\,\AA) and \ion{Ca}{I} (at 6572.779\,\AA) observed in the visual \citep[from][]{kamiV4332_2010}. The narrow absorption dips in the profile of \ion{K}{I} are mainly interstellar lines. The correction to heliocentric velocity is --12.144\,\kms.}\label{fig-profiles}
\end{figure}

SiO 8--7 is the strongest observed spectral feature. Its profile representing the entire emission region has a FWHM of 146\,\kms, is fairly symmetric, and well reproduced by a Voigt profile. In order to illustrate the symmetry of the profile, we compare it to a synthetic profile in Fig.\,\ref{fig-profiles}. CO $J$=3--2 is another strong line. The SiO and CO emission integrated over the same region have slightly different profiles; overall, the CO line with a FWHM of 210\,\kms\ is wider and rounder. The difference may be caused by different spatial distributions of CO and SiO within the environment of varying excitation conditions as the upper level energies of their transitions are of 33.19\,K (CO) and 75.02\,K (SiO). Additionally, the CO line displays a broad parabolic ``pedestal" which interpolated over the entire profile would have a full width of nearly 500\,\kms (Fig.\,\ref{fig-profiles}). 

Lines of SO$_2$, SO, and AlO are usually blending with each other so that their profiles cannot be studied in detail. Although a weak broad component is present in the spectra of these species, e.g. in SO$_2$ in Fig.\,\ref{fig-profiles}, their profiles are dominated by a very narrow core of a width of $\sim$50\,\kms, much narrower than that of SiO and CO. This narrow component is likely linked to the smaller spatial extent of emission of these species (Sect.\,\ref{sec-v4332-sizes}). In the broader component, the red wing is slightly stronger. 
 	
In Fig.\,\ref{fig-profiles}, the submm lines are compared to a sample of atomic lines observed in the visual spectrum of V4332\,Sgr in 2005 and 2009 in  \citep{kamiV4332_2010,tyl2015}. The \ion{Ca}{I} $\lambda$4226 line (top panel) is very saturated and among the broadest atomic lines in this source. At half maximum, it is broader than the CO profile, but both lines have comparable widths at the base of the profile. A less saturated visual line of \ion{K}{I} $\lambda$7699, is narrower than the line of CO and its shape is similar to that of SiO. The profiles of SO lines are similar to the narrowest atomic lines, as illustrated with the $\lambda$6572 line of \ion{Ca}{I} (cf. middle and bottom panel). These similarities indicate that the neutral atomic gas seen in visual spectra and the molecular gas revealed by ALMA are located within the same kinematic structure. The variations in line profiles reflect varying excitation conditions and abundance patters of the different species. 

Observed during the 1994 outburst \citep{martini}, the H$\alpha$ emission line had a FWHM of about 200\,\kms, comparable to that of the CO 3--2 profile from 2016. The molecular gas was very likely ejected in 1994. Assuming that the velocity dispersion in tangential motions is the same as that measured in radial velocities, with a FWHM/2 of about 105\,\kms, and given the $e$-folding radius of the CO emission of 116\,mas to which the remnant expanded in 22\,yr\footnote{We ignored the broad line pedestal that indicates slightly higher velocities because it is not the most spatially extended component.}, the object is at a distance of $\sim$4.2\,kpc. This value is slightly smaller than the lower limit of 5.5\,kpc derived by \citet{tyl2015} from velocities of the interstellar absorption lines but is consistent within the large uncertainties of both estimates. We assume the value of 5\,kpc as the representative distance to V4332\,Sgr. At this distance, the largest radius of CO emission translates to 580\,AU. 

\citet{tyl2015} derived a heliocentric radial velocity of V4332\,Sgr of --75.1$\pm$3.1 ( or --63.0 \kms\ LSR) while \citet{kamiV4332_2010} derived --65$\pm$7\,\kms\ (--53\,\kms\ LSR). These estimates are based on optical observations of mainly atomic features. The sharp and unblended profile of SiO $J$=8--9 is centered at $V_c$=--54.11$\pm$0.96 \kms\ relative to the LSR (or at --66.25\,\kms\ heliocentric). A very similar value was derived from the CO profile. The optical thickness in the SiO and CO line is unknown but a large portion of the observed emission may arise from optically thick regions (see Sect.\,\ref{sec-masses}). The weaker and narrower lines of SO$_2$ may be still optically thinner and the cumulative Gaussian fit to all detected lines implies a radial LSR velocity of --62.3$\pm$2.5\,\kms, slightly more blue-shifted than that of the stronger lines but the asymmetric broad wings of SO$_2$ might distort this measurement. 

\subsection{Spatio-kinematic structure}\label{sec-kinema}
\begin{figure*}[!h]
\sidecaption
\includegraphics[width=6cm,trim={10 10 50 10},clip]{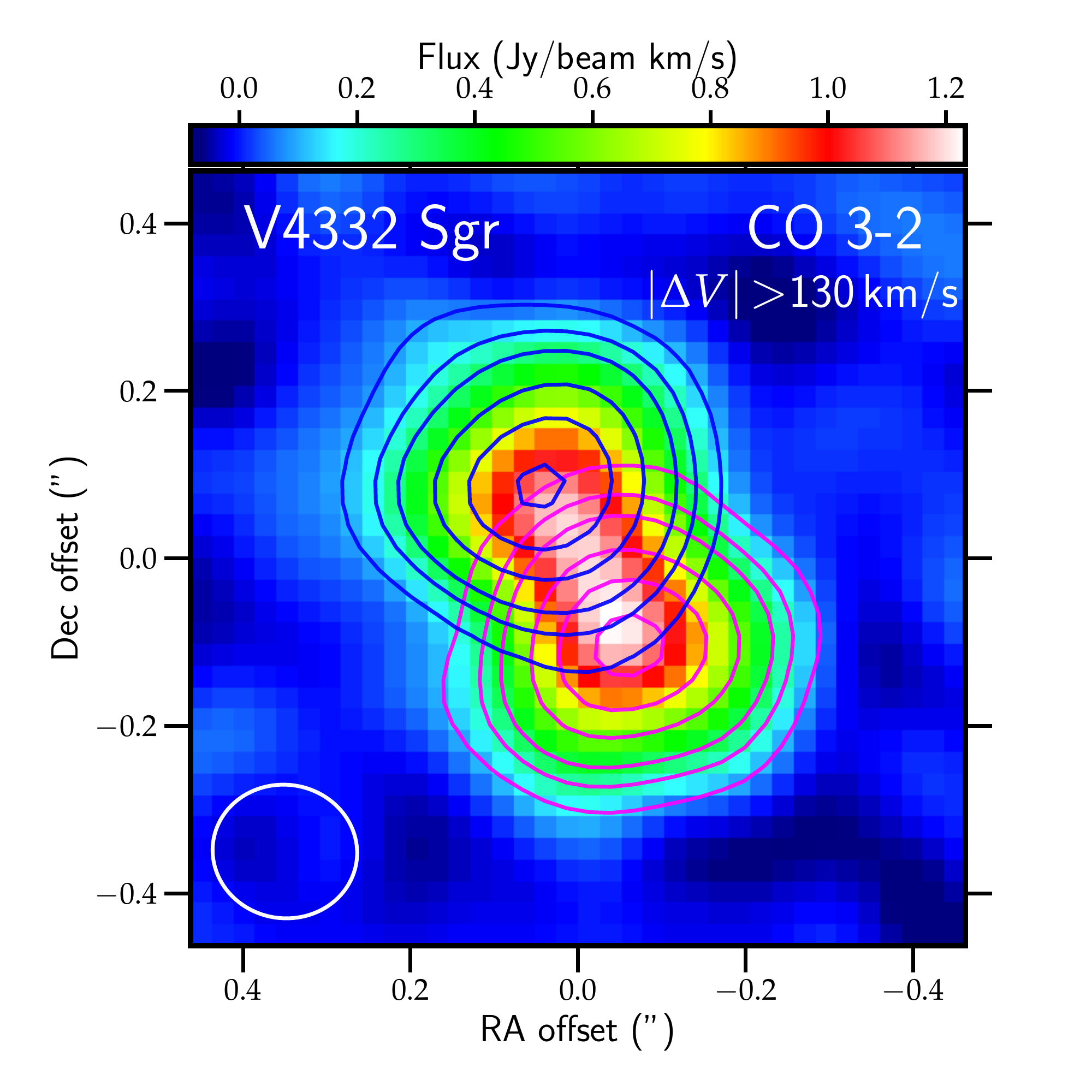}
\includegraphics[width=6cm,trim={10 10 50 10},clip]{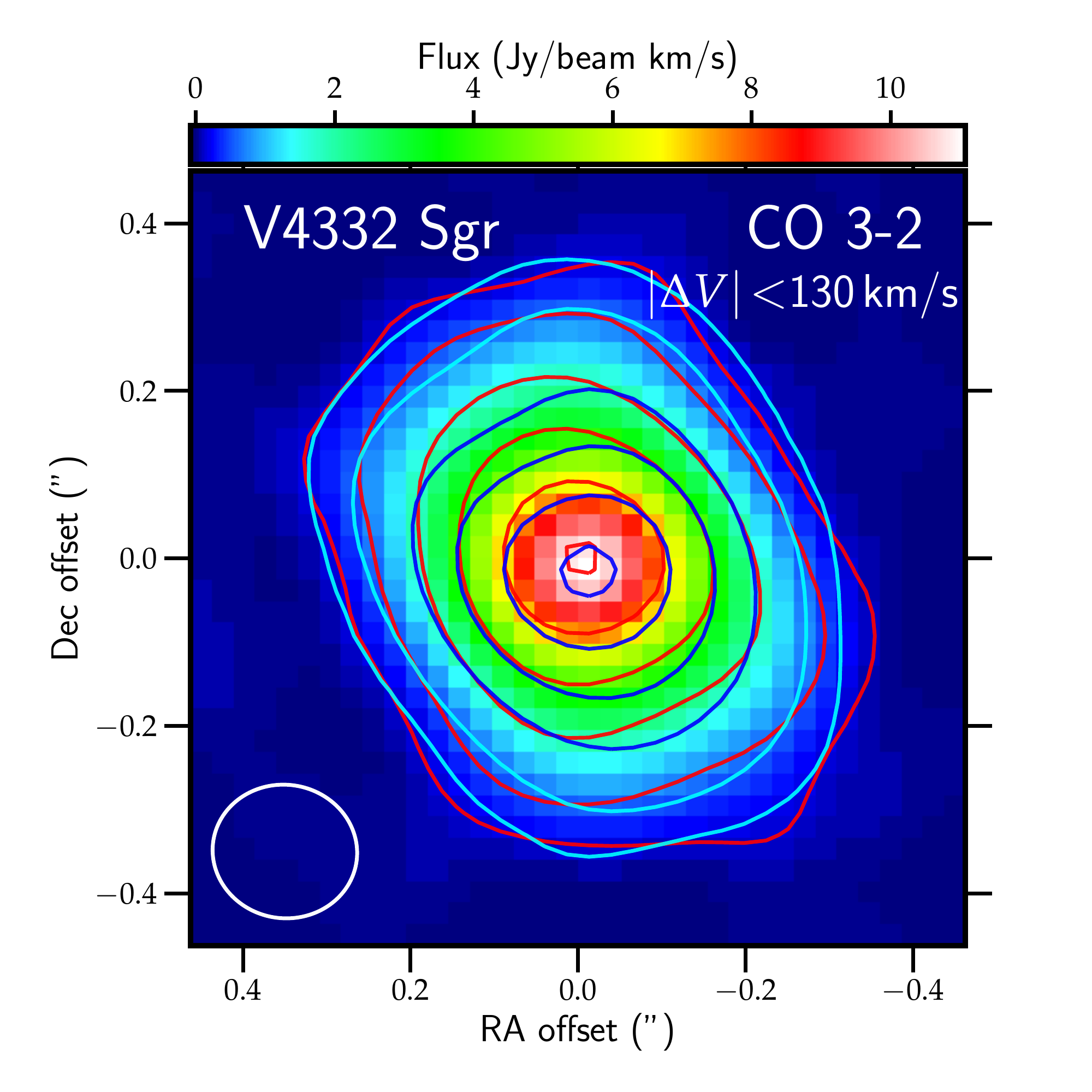}
\caption{Maps of CO emission in V4332\,Sgr. {\bf Left:} The high velocity outflow with $|\Delta V|>130$\,\kms. The blue and magenta countours show the blue- and red-shifted components, respectively, and the image shows the summary flux. The coutours are drawn at levels of [5, 10, 15, 25, 35, 45] times the respective rms. {\bf Right:} The intense CO core emission at $|\Delta V|<130$\,\kms. The countours are drawn at [5, 15, 45, 90, 160, 235] times the rms; red and blue/cyan contours represent emission that is respectivy red- and blue-shifted (relative to $V_{\rm LSR}$=--54\,\kms). The background image shows the cumulative flux of both components.}\label{fig-co-comp}
\end{figure*}

The ALMA maps readily reveal a rather complex spatio-kinematic structure of the merger remnant in V4332\,Sgr. Overall, blue-shifted emission (w.r.t. to the central velocity, $V_c$) forms the northern, N, lobe and red-shifted one defines the southern, S, lobe. Surprisingly, the gas component at most extreme velocities, i.e. with $|\Delta V|$=$|V-V_c|>$130\,\kms\ and marked in Fig.\,\ref{fig-profiles} as the ``parabolic'' pedestal, is not the most extended emission. Maps corresponding to these most extreme velocities (see Fig.\,\ref{fig-co-comp}) show beam-smeared emission peaking $\sim$90\,mas off and extending to about 160\,mas north and south (along a PA of 31\degr) from the overall center of CO emission. The location of this high-velocity gas can also be identified in Fig.\,\ref{fig-residuals} as the most negative residuals in the CO map. The redshifted lobe has a peak flux that is 1.03 times stronger than that on the blueshifted side. This slight asymmetry is also seen in the spectral profile in Fig.\,\ref{fig-profiles}. 

The line core emission of CO with $|\Delta V|$<130\,\kms\ is spatially more extended than the high-velocity component, as shown in Fig.\,\ref{fig-co-comp}. The redshifted and blueshifted halves of this core emission show similar spatial distributions but at a close inspection the redshifted emission appears slightly stronger and more extended in the N lobe (and the blue component in the S lobe), i.e. in opposite to what is seen in the high-velocity component. The maximum radial extent of the (beam-smeared) CO emission is $\sim$370\,mas.

\begin{figure*}[!h]
\includegraphics[width=0.33\textwidth,trim={0 0 0 0}]{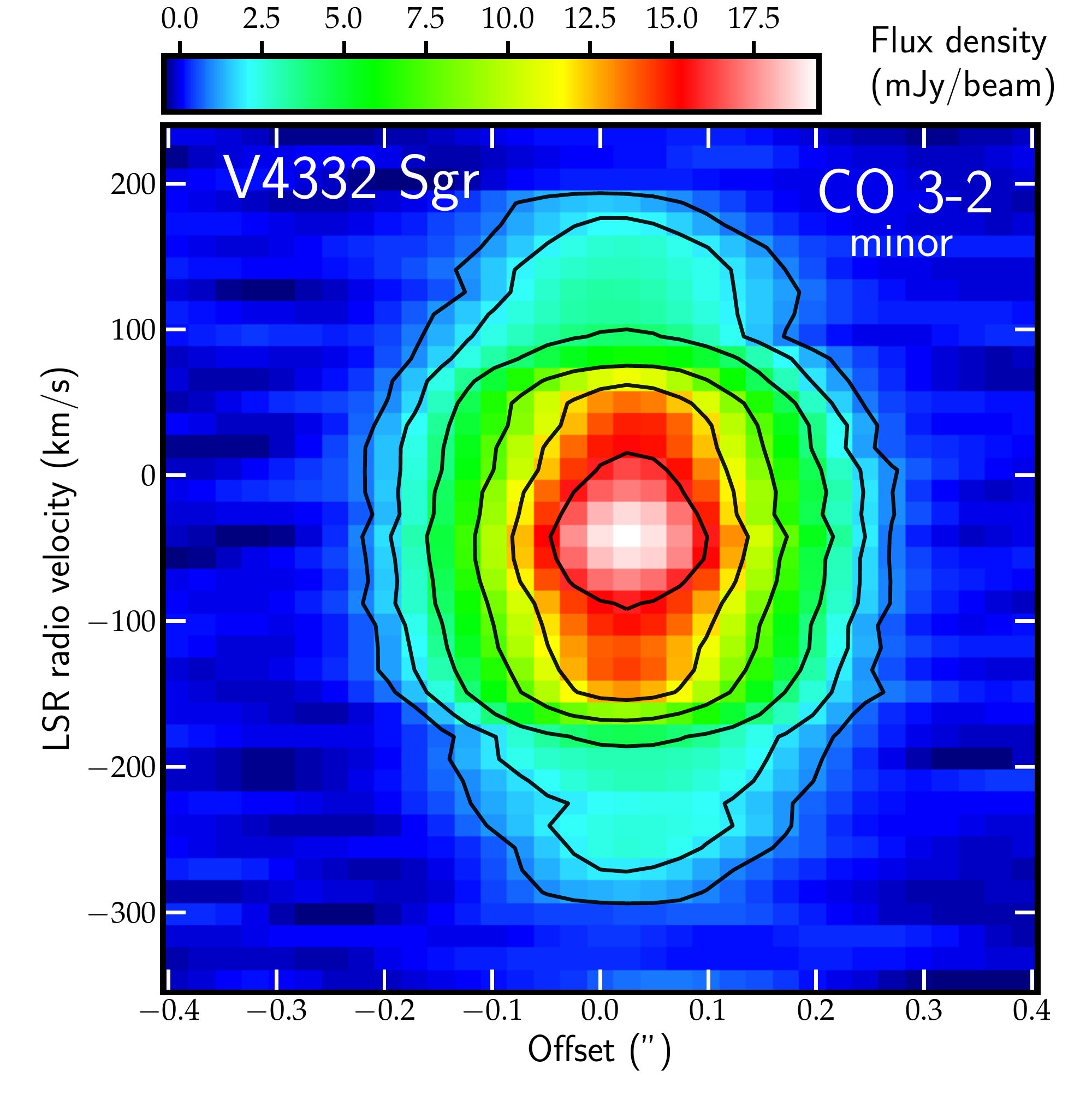}
\includegraphics[width=0.33\textwidth,trim={0 0 0 0}]{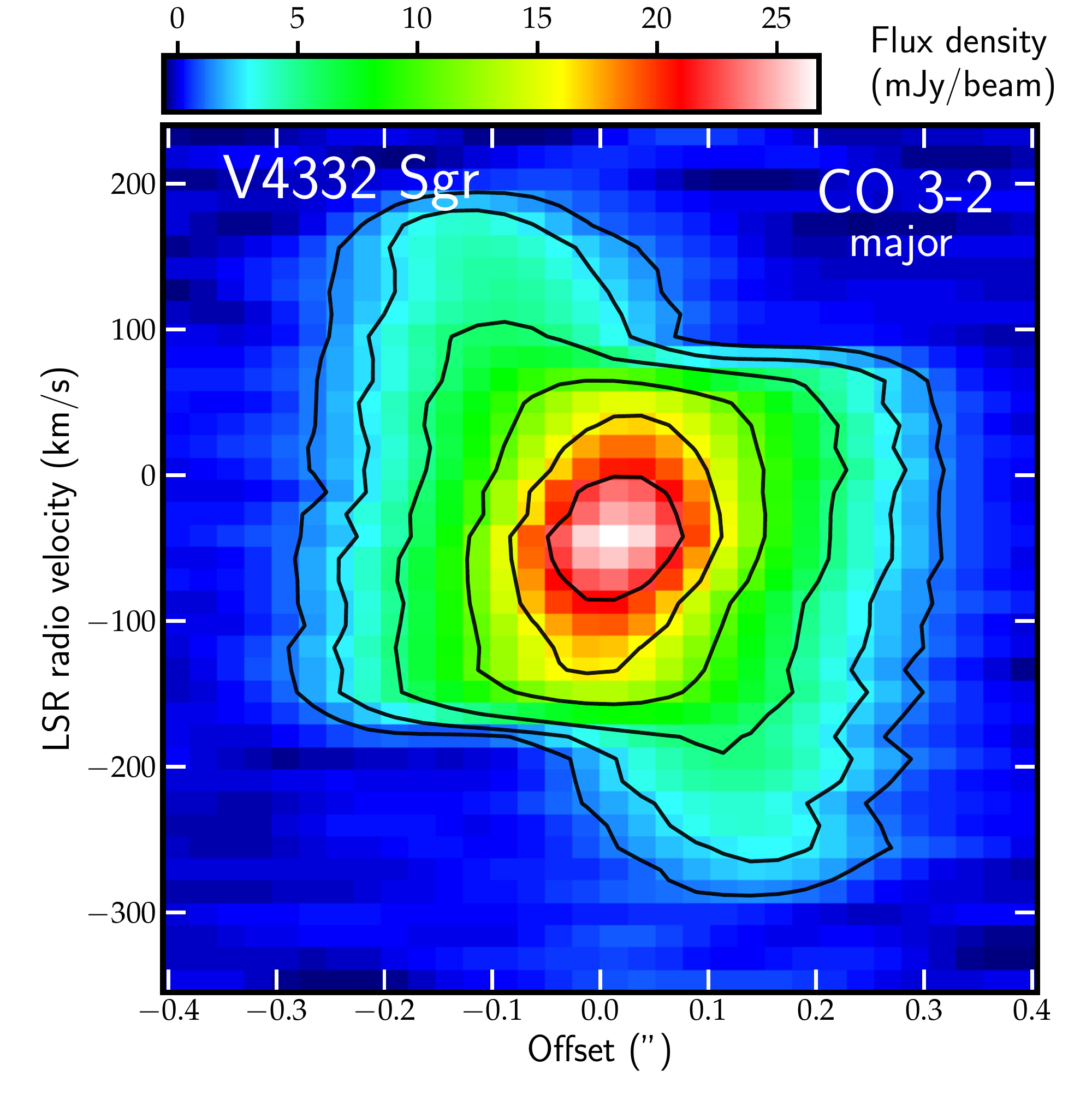}
\includegraphics[angle=0,width=0.33\textwidth, trim={0 0 0 0},clip]{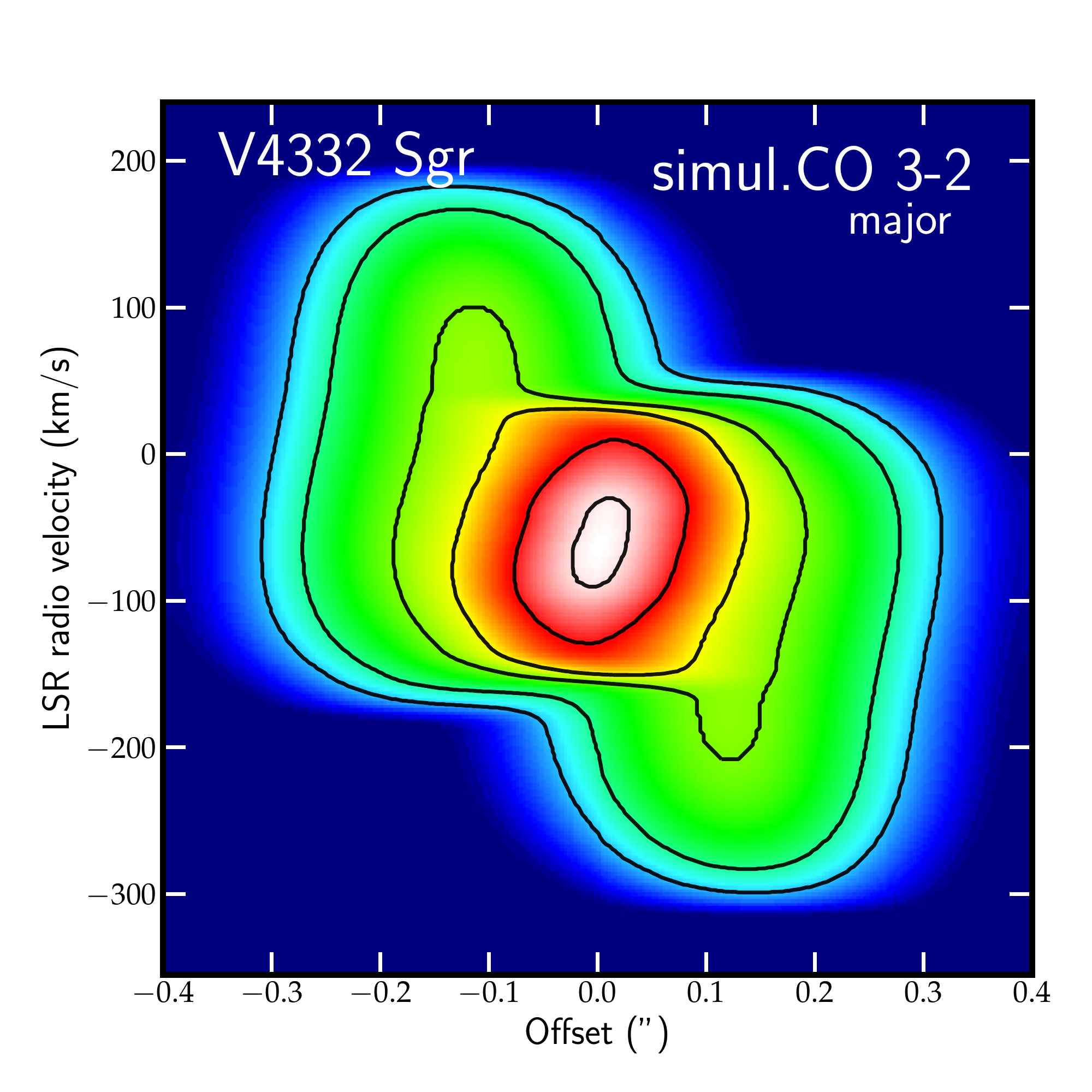}
\caption{Position-velocity diagrams of V4332\,Sgr. The diagrams represent the entire emission of CO (image and contours) along the minor (left panel) and major (middle panel) axis of the region. The left and middle panel are based on ALMA observations while the right panel shows a sample diagram extracted along the longer axis of the CO emission region simulated in Shape \citep{shape}. The contours are drawn at 5, 10, 20, 40, 60, and 80\% of the observed emission peak. The contours of the Shape simulation are drawn at 10, 20, 45, 60, 80, 97\% of the peak.}\label{fig-PV}
\end{figure*}

The complex structure of the outflow is well illustrated in the position-velocity (PV) diagram extracted along the longer axis of the CO emission region and shown in middle panel of Fig.\,\ref{fig-PV}. There, the strong central-velocity component has a velocity gradient. The position-velocity diagram extracted along the minor axis, shown in the left panel of Fig.\,\ref{fig-PV}, does not display any velocity  gradients. This indicates there is no rotation or accelerating outflow in the direction perpendicular to the main outflow axis. It is again apparent that the low-velocity component, at LSR velocities from --190 to 90\,\kms, is more extended than the component at extreme velocities.

\subsection{A 3-dimensional model of the CO emitter}
\begin{figure*}[!h]\sidecaption
\fbox{\includegraphics[width=7.46cm, trim={0 0 0 0},clip]{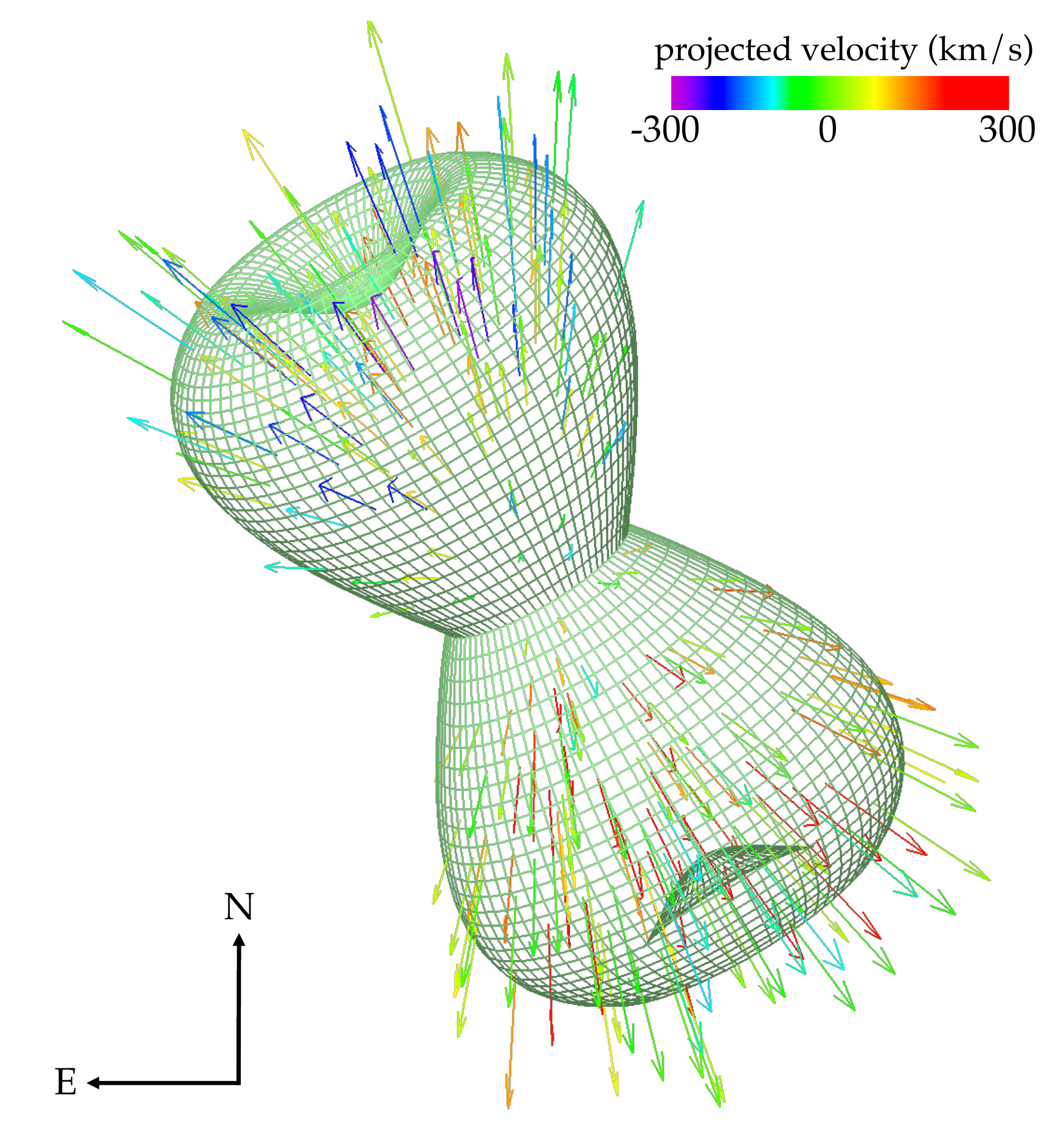}}
\fbox{\includegraphics[angle=0,width=4.54cm, trim={0 0 0 0},clip]{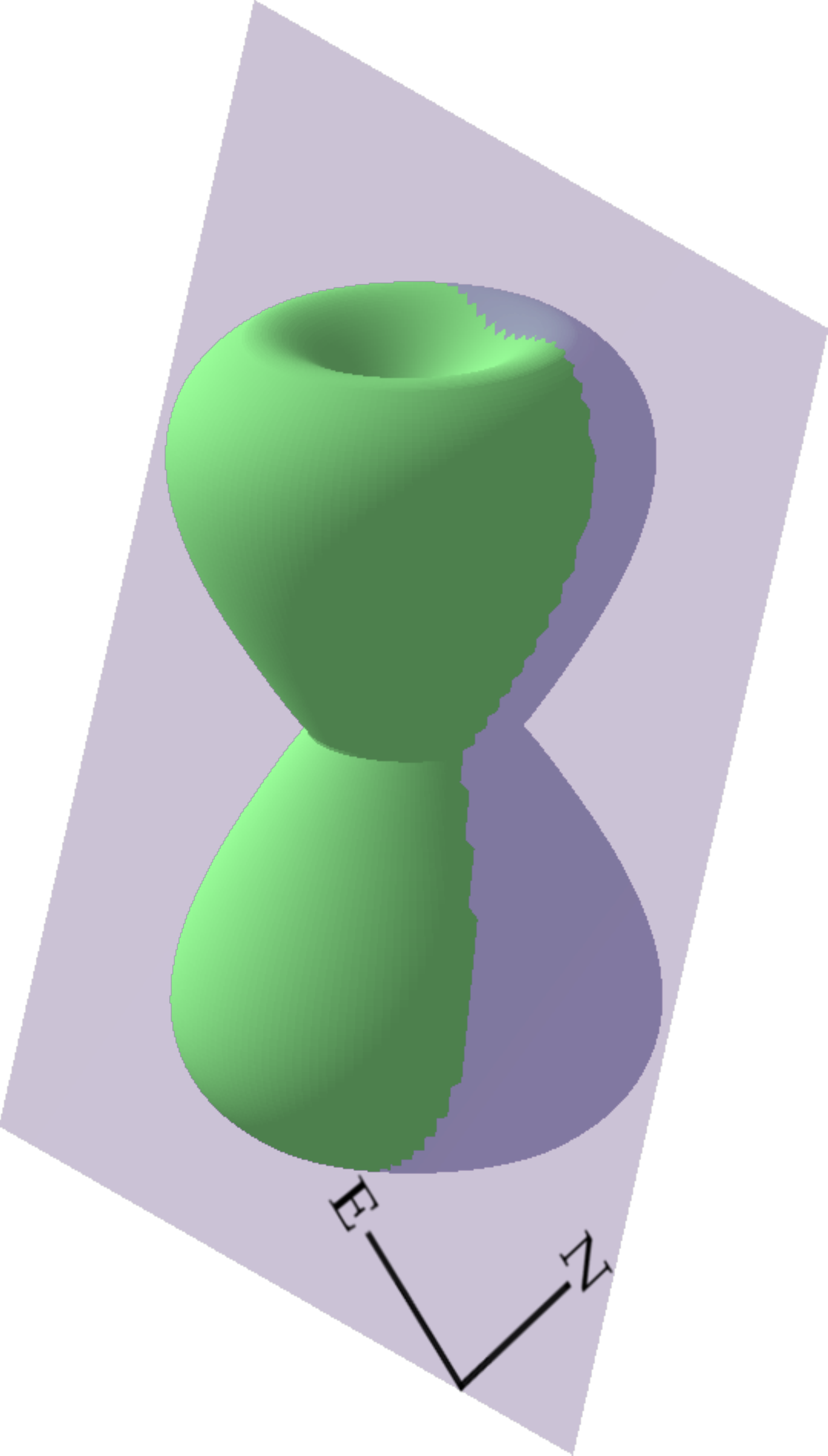}}
\caption{Model of the molecular outflow in V4332\,Sgr. {\bf: Left:} Observer's view on the ouflow at $i$=13\degr. The arrows represent the velocity field ($\vec{\varv} \propto \vec{r}$) within the volume of gas filling the bipolar structure. Their length is proportional to the velocity and their colors represent the projected velocity with color-coding indicated by the colorbar. {\bf Right:} The same structure shown from a side view. The purple plane represents the plane of the sky. The observer is located approximetely to the left and the purple-shaded part of the bipolar structure is the far side for the observer.}\label{fig-Shape}
\end{figure*}

The complex spatio-kinematic characteristics of CO emission in V4332\,Sgr can be interpreted as arising from a biconical outflow seen at a very small inclination angle, in particular much smaller than its opening angle. (We define the inclination as an angle between the long axis of the biconical structure with respect to the plane of the sky and counted positive from north towards the observer.) In an effort to test this interpretation, we constructed a 3-dimensional model of the structure in Shape \citep{shape,shape2,shape3}, including basic radiative transfer of CO utilizing the {\tt shapemol} module \citep{shapemol}. The model was optimized by referring to the total intensity and channel maps, PV diagrams, and spectral profiles of the CO 3--2 emission. The final spatial model is shown in Fig.\,\ref{fig-Shape}.

As the velocity field, we adopted homologous expansion, where the velocity is proportional to the distance from the source, as it is the simplest field that reproduces the observations relatively well. In the basic model, the CO emission is dominated by gas filling a bipolar structure with a wide opening angle of $\sim$60\degr, a de-projected radius corresponding to 0\farcs28, and low inclination angle of 0\degr $< i \lesssim$30\degr. At these inclinations, the maximum unprojected velocities that we implemented were $\sim$420\,\kms. Models in which the bipolar structure is a hollow shell do not reproduce observations well, indicating that the intense CO emission arises from nearly the entire volume of the structure. The wide opening angle coupled with the linear velocity field results in the relatively wide velocity dispersion observed in the core of CO emission profile. The extreme radial velocities with $\Delta V > 130$\,\kms\ are related to the inclination of the structure: the extremely blueshifted northern component arises in the nearest parts of the inclined structure while the extreme red component is produced in the far southern side. Other main kinematic characteristics of CO observed in V4332\,Sgr are well reproduced by this model, as illustrated with the simulated PV diagram in Fig.\,\ref{fig-PV}.

The Shape simulations also indicate that the opacity of the extended emission of the CO 3--2 line is low, i.e. $\tau \ll$1. Versions of the model with high CO opacities display characteristic asymmetry in velocity, clearly distinguishable in PV diagrams, which is not seen in the ALMA observations of V4332\,Sgr.

Even our simplest model with a single geometrical structure required rather complex density profiles. There is strong degeneracy between the velocity field and the density distribution which could not be broken with the current observations. There are certainly more complex models that can explain the observations equally well or better than our simple hourglass model. By performing extra simulations we find, however, that there are no observational features that could be unambiguously identified as a torus or a rotating disk-like structure in the plane parallel to the major axis of the bipolar outflow. A slightly better match with observations is achieved after adding to the basic hourglass outflow a pair of collimated jets moving at velocities significantly higher than the bulk of the wide-angle outflow. These jets are embedded in the central part of remnant and extend to a radius corresponding to 80\,mas. Their flux contribution to the total emission is however very modest and their kinematic signature is subtle so the presence of the jets is only a tentative result. 

\subsection{Column densities, excitation, and the mass}\label{sec-cassis-v4332}
In an effort to derive basic physical parameters of the molecular gas in V4332\,Sgr, we performed an excitation analysis of the ALMA spectra in CASSIS. We used procedures that implement local thermodynamic equilibrium (LTE) conditions and allow CASSIS to search for best set of physical parameters (line width and positions, an excitation temperature, and a column density) by performing a $\chi^2$ minimization procedure. The line profiles are represented by Gaussian profiles corrected for opacity effects. Source sizes were fixed at values derived in Sect.\,\ref{sec-v4332-sizes}. We analyze here a spectrum extracted for the central pixel which has the highest S/N for most of the species and is representative of gas within the central beam of FWHM of 165\,mas, i.e. does not represent the extended component seen in lines of CO and SiO. The simulated LTE spectra are shown in Fig.\,\ref{fig-cassisModel-V4332}. 

A full excitation analysis could only be performed for SO$_2$ which is traced in multiple transitions arising from levels in a wide range of energies above the ground, $E_u$=48.5--320.9\,K. The simulations yield an SO$_2$ excitation temperature $T_{\rm ex}$=93.5 (68.5--136.0) K and a column density of 5.7 (4.8--6.6) $\times 10^{16}$\,cm$^{-2}$ (values in brackets are 3$\sigma$ ranges). From the absence of emission in the $^{34}$SO$_2$ isotopologue, whose lines were covered by the spectrum, we derive an upper limit on its column density of $\sim$2.6$\times 10^{15}$\,cm$^{-2}$. It is consistent with the solar isotopic abundance ratio, $^{34}$S/$^{32}$S=1/22.35.   

The SO central line velocity, the line width, and the angular size of the emission region are consistent with these for SO$_2$. We therefore calculated the SO column density for the same $T_{\rm ex}$ as that of SO$_2$ and found $N$=2.7$\times 10^{16}$\,cm$^{-2}$. 

As only one rotational transition was covered of each AlO and AlOH, their excitation temperature cannot be determined and we assumed the same $T_{\rm ex}$ for the Al-bearing species as that of SO$_2$. However, the physical co-location of Al-bearing species and SO$_2$ is not certain and their emission may be excited under different conditions. The line intensities then yield a column density of 1.3$\times 10^{15}$\,cm$^{-2}$ for AlO. The abundance of AlOH (if truly present in the spectrum) is $\sim$2.7 times higher that that of AlO. Electronic bands of AlO have been observed in the visual spectra of V4332\,Sgr for over a decade but owing to the mechanism in which the visual bands arise, i.e. resonant scattering, it is not straightforward to constrain the quantity of the circumstellar AlO gas. The excitation temperature 93\,K that we adopted for AlO is slightly lower but within uncertainties consistent with a rotational temperature of $\sim$120$\pm$20\,K constrained from 2009 spectra of optical electronic bands of other metal oxides, including TiO, VO, and ScO. Their emission is produced in the same radiative process as that in AlO bands \citep{kamiV4332_2010}. We thus believe, that the AlO column density and temperature are relatively well constrained.

The emission in SiO and CO indicates a presence of several spatio-kinematic gas components which may be characterized by different temperatures. Some of the components are extended and therefore their excitation temperature is very likely lower than that of SO$_2$. Nevertheless, assuming the temperature of 93\,K for CO and SiO and adopting line widths of 170 and 100\,\kms, respectively, we derive column densities of 6.3$\times 10^{18}$\,cm$^{-2}$ for CO and 2.2$\times 10^{16}$\,cm$^{-2}$ for SiO. The resulting lines are only moderately saturated at these parameters. At lower excitation temperatures, the simulated profiles saturate considerably and it is impossible to satisfactorily reproduce the observed peak line intensities with a single Gaussian component. Such a saturated profile, however, fits very well to the main part of the CO line profile. Also, to explain the CO and SiO emission at lower excitation temperatures, the intrinsic line widths need to be much lower and the column densities much higher than these quoted above. To illustrate this effect, let us consider the following. For saturated emission homogeneously and entirely filling the beam, the brightness temperature of the line is expected to approach the true excitation temperature of the gas at opacity near $\tau$=1. The Rayleigh-Jeans brightness temperatures in the peaks of the CO and SiO lines are of about 23 and 34\,K. These brightness temperatures set rough lower limits on $T_{\rm ex}$ and are much lower than the temperature derived from the SO$_2$ observations. In Fig.\,\ref{fig-cassisModel-V4332}, we show sample single-component profiles of the two lines simulated at $T_{\rm ex}$=34\,K. This temperature requires unrealistically high column densities to approach the observed line intensities. This perhaps indicates that even the densest parts of the remnant are actually optically thin in CO 3--2 (or that the assumptions of a single isothermal gas component in LTE is inadequate for this object). While these considerations are based on the spectrum representing the brightest part of the remnant, Shape modelling presented in Sect.\,\ref{fig-Shape} shows that also the extended molecular emission is optically thin.

We use the constraints on the CO column density at $T_{\rm ex}$=93\,K to derive the total mass of the material. Assuming the distance of $d$=5\,kpc (Sect.\,\ref{sec-profiles-distance}), a source size of 232$\times$165\,mas, and CO abundance relative to H$_2$ of $3 \times 10^{-4}$, we get the total mass $\sim$0.01\,M$_{\sun}$ (including He at solar abundance). 

\section{V1309 Sco}\label{sec-v1309}
\subsection{ALMA observations}\label{sec-obs-v1309}
\paragraph{Bands 3,4,6}First ALMA observations of V1309\,Sco were obtained in Jun.--Jul. 2015 (PI: B. McCollum) in Bands 3, 4, and 6. The observations in Bands 3 and 4 did not yield any detection, neither in lines nor in continuum,  and reached  continuum sensitivities of 87.0\,$\mu$Jy per beam of FWHM of 0\farcs55 and 46.3\,$\mu$Jy per beam of FWHM of 0\farcs35, respectively, each in a cumulative bandwidth of 7.6\,GHz. The Band\,6 observations also failed to detect continuum emission\footnote{Although McCollum et al. claim in \url{www.astro.caltech.edu/~sma/dynamicirsky.pdf}, page 40, that continuum was detected, we find that all flux in this band comes from molecular emission, chiefly from CO $J$=2--1.} (at an rms of 89.0\,$\mu$Jy per beam of FWHM of 0\farcs23) but resulted in a detection of molecular emission which we include in our analysis below. The Band\,6 data were taken on 6 July 2015 with 39 antennas and included observations of a bandpass calibrator, J1733-1304, and two gain calibrators, J1744-3116 and J1743-3058. Titan was observed for absolute flux calibration. We calibrated the data with the CASA's pipeline version 4.2.2 and used the default calibration script. The script was modified to include extra flagging of bad-performance antennae and of channels corresponding to CO $J$=2--1 absorption in the phase calibrator J1744-3116. The detected molecular emission in V1309\,Sco is rather weak and self-calibration was not feasible. The projected baselines used in the observations ranged from 43.3\,m to 1.6\,km. To increase the sensitivity, we imaged the data with natural weighting of visibilities which resulted in a nominal beam size of 252$\times$219\,mas. The spectra were registered at a resolution of 15.625\,MHz (or $\sim$20\,\kms) and covered 230.1--233.9 and 246.1--249.9\,GHz.

\paragraph{Band 7}
Our ALMA observations of V1309\,Sco in Band\,7 (PI: T. Kami\'nski) had about 10 times longer integration time than the earlier observations at longer wavelengths. The spectral resolution and frequency coverage in this band were very similar to that in ALMA observations of V4332\,Sgr (Sect.\,\ref{sec-obs-v4332}). The same observational setup was executed on 17 January, 26 March, and 9 April 2016, and  used 39, 42, and 43 antennas, respectively, resulting in a combination of 58 different antenna positions. The combined projected baselines ranged from 14.8 to 455.8\,m giving a nominal beam FWHM of 408$\times$352\,mas at uniform weighting and largest recoverable scales of 7\farcs2, i.e. much larger than any emission region observed in the field of view. The quasar J1924-2914 was observed to calibrate the bandpass. Flux calibration was performed by observing Titan on the first date and J1733-1304 on the two later dates. The complex gains were determined by observations of J1744-3116 in the two earlier dates and J1733-3722 in the last execution. The calibration was performed with CASA's pipeline version 4.5.3 and the default calibration script was used. We attempted self-calibration of the V1309\,Sco visibilities by referring to the peak of the strongest emission feature, CO $J$=3--2, but it did not improve the S/N, which is of 74 for the CO peak at a 15\,\kms\ resolution.

Although V1309\,Sco was not detected in Band\,7 continuum, ALMA detected an unrelated continuum source at RA=17$^h$57$^m$32\fs6768 and Dec=--30\degr43\arcsec14\farcs157 (J2000.0), i.e. 5\farcs4 south-west from V1309\,Sco. The source has a flux of 1.59$\pm$0.06\,mJy and a size of 715\,mas. It is not listed in any of the CDS/Aladin catalogs (as of Dec. 2017) but a {\it Hubble} visual image in the $F673N$ filter \citep[][]{v1309sed} shows a possible optical counterpart. 

\subsection{Identification of spectral features}
\begin{figure*}[!h]
\includegraphics[angle=270,width=0.28\textwidth]{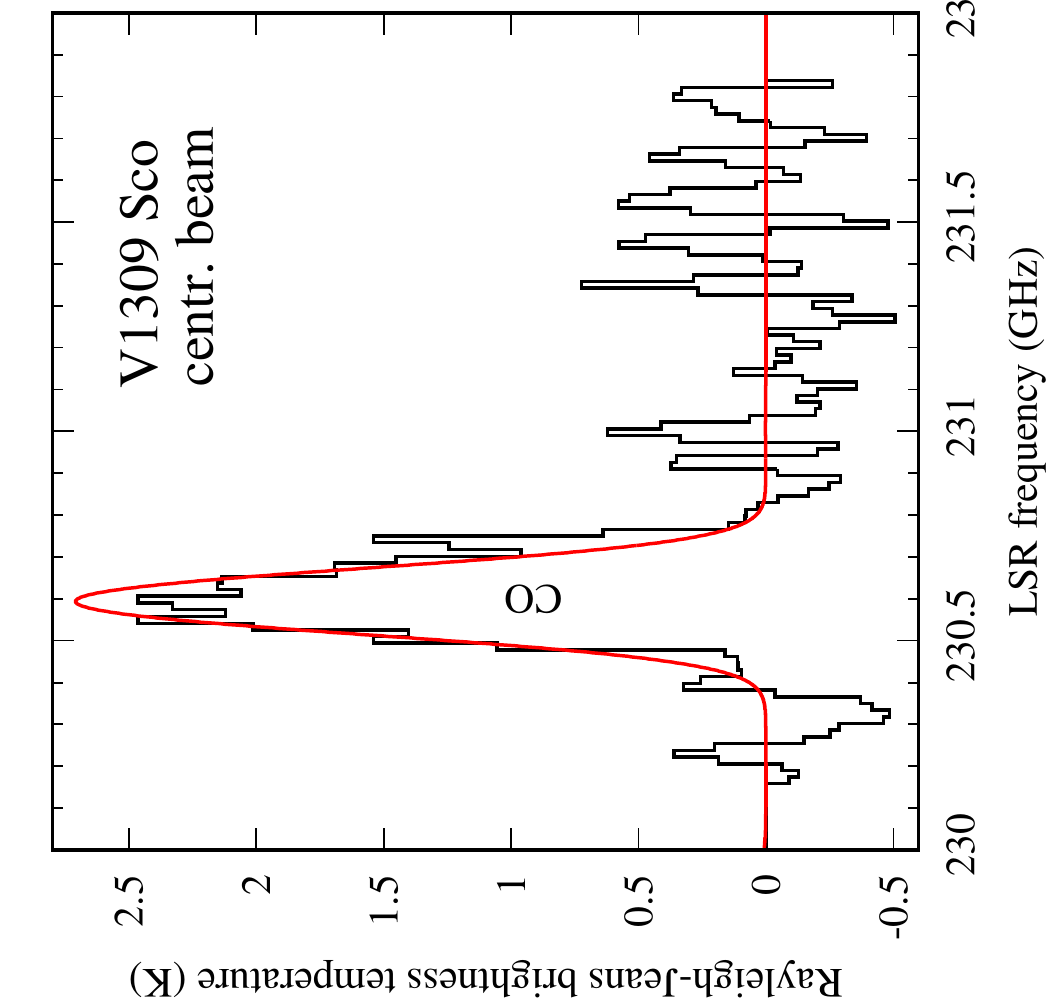}
\includegraphics[angle=270,width=0.7\textwidth]{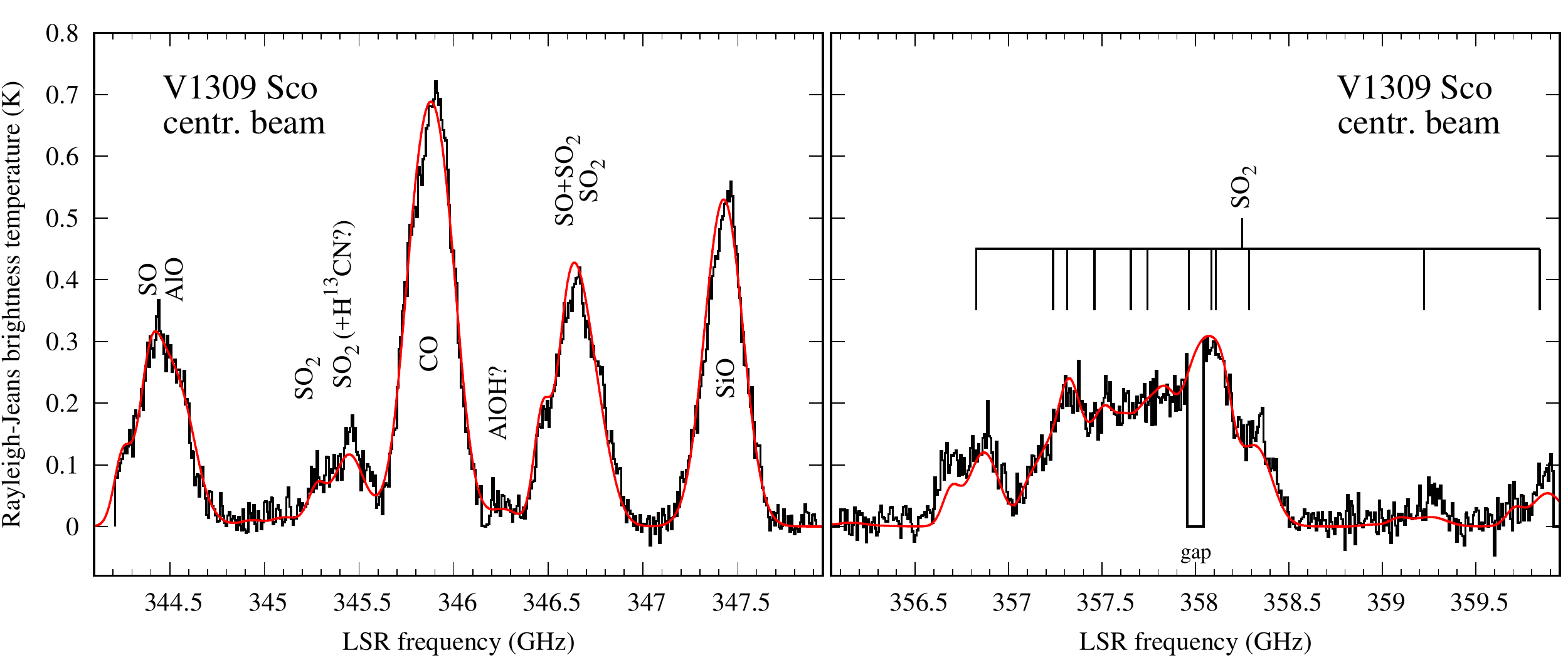}
\caption{ALMA spectra of V1309\,Sco (black) and their simulations (red). The observed spectrum was extracted for the central beam of the molecular region and is displayed in the Rayleigh-Jeans brightness temperature scale. In the CASSIS simulation, we adopted excitation temperatures of 36\,K for CO and of 80\,K for most other species (see text). SO, SO$_2$, and AlO were modelled with two velocity components.}\label{fig-cassisModel-V1309}
\end{figure*}

The identification procedure for the lines was analogous to that described for ALMA spectra of V4332\,Sgr (Sect.\,\ref{sec-ident_v4332}) and the list of lines detected in V1309\,Sco, summarized in Table\,\ref{tab-lines}, contains essentially the same transitions as these found in V4332\,Sgr. The 14\,yr younger red nova V1309\,Sco has broader and more complex line profiles. Line blending is more severe, especially for the series of SO$_2$ transitions above 356.5\,GHz and for the AlO and SO blend near 344.45\,GHz. In the Band 3, 4, 6 spectra available exclusively for V1309\,Sco, only one other line is detected, i.e. CO $J$=2--1. The spectra with assigned lines are shown in Fig.\,\ref{fig-cassisModel-V1309}.

\subsection{Location and size}
The Band\,6 data covering the CO 2--1 line have the highest angular resolution among all the data collected so far for V1309\,Sco but have also a very modest S/N (of $\sim$10 in the integrated intensity map) which limits the usability of the data. A Gaussian fit to the map of this transition yields a beam-deconvolved size of (254$\pm$32)$\times$(218$\pm$36)\,mas. 

The numerous transitions in Band\,7 observed at a nearly twice lower angular resolution but at a much higher S/N allow for more precise position and size constraints. The absolute position derived from a Gaussian fit to cumulative flux in all lines is RA=17$^h$57$^m$32\fs9349 ($\pm$2.1\,mas) and Dec=--30\degr43\arcmin09\farcs9421 ($\pm$1.8\,mas, J2000.0). Central location of emission regions corresponding to individual features are within uncertainties consistent with this measurement. The size of the region representing all emission lines is (170$\pm$ 21)$\times$(105$\pm$46)\,mas, but there are size differences between the different emission features. In contrast to what we measured in V4332\,Sgr, in V1309\,Sco the emission region of CO 3--2 with a beam-deconvolved size of (144$\pm$24)$\times$(114$\pm$39)\,mas is not the most extended one. The measured size of SiO is slightly larger with dimensions of (164$\pm$25)$\times$(136$\pm$36) mas. The major axis of SO$_2$ emission is even more extended with a FWHM of 176$\pm$26\,mas and the emission region is narrowest among all these that were measured,  with the minor axis of 84$\pm$68\,mas. The longer axes of emission regions corresponding to the different spectral features are all well aligned at a PA of 151\degr$\pm$18\degr. Because the synthesized beam is a few times larger than the (beam-deconvoled) Gaussian sizes, no substructure is recognizable to the eye in the emission maps. 

\subsection{Kinematics of V1309\,Sco}
\begin{figure}[!h]
\includegraphics[width=0.45\textwidth]{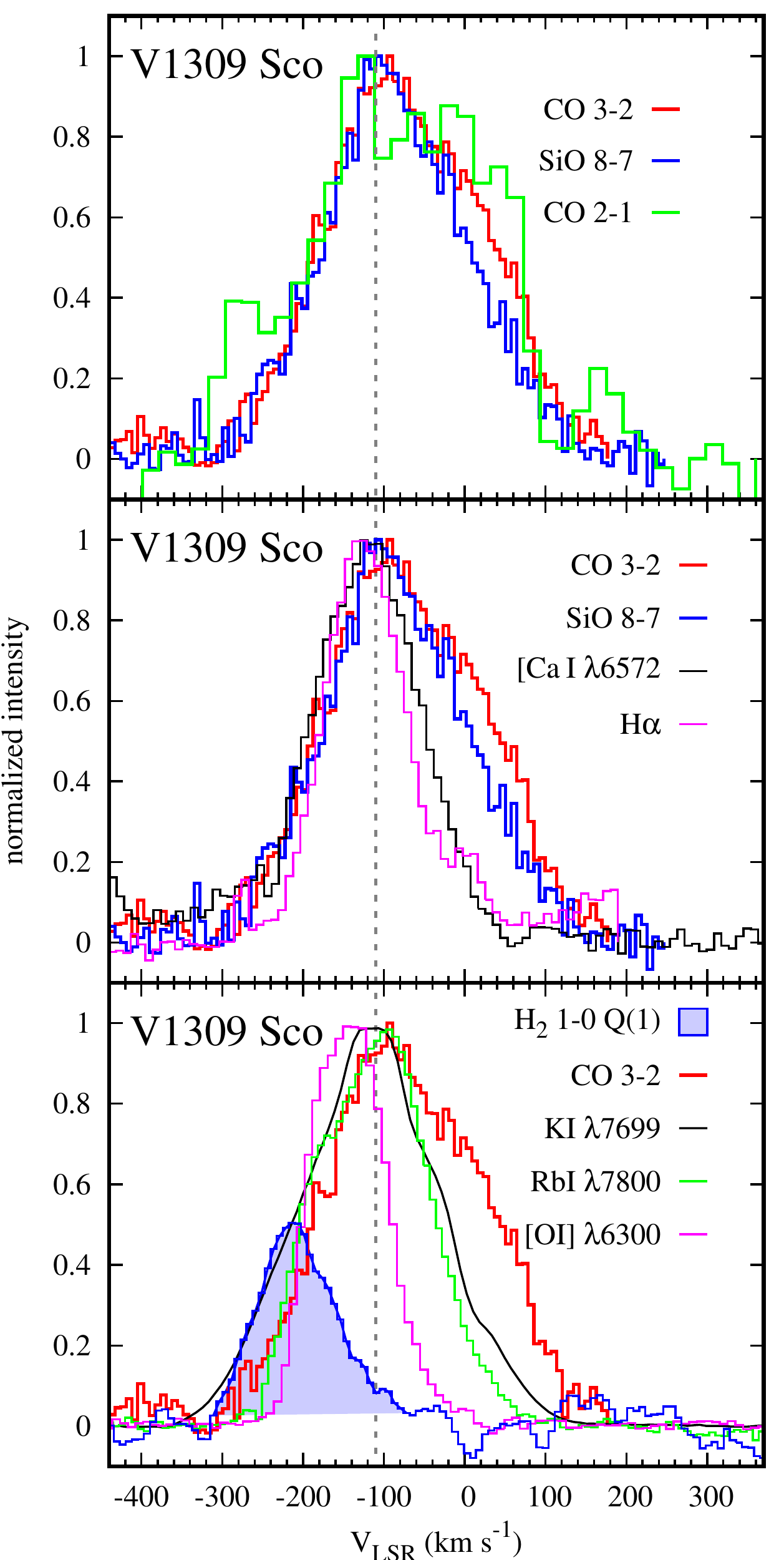}
\caption{Sample emission profiles in V1309\,Sco. They represent the entire emission region. {\bf Top:} The profiles of rotational emission of SiO 8--7 and CO 3--2 observed in Jan.--Apr. 2016; and of CO 2--1 observed in July 2015. {\bf Middle and bottom:} The molecular profiles observed in 2016 are compared to sample profiles observed in May--Jul. 2016 in the visual. The profile of \ion{K}{I} is affected by narrow telluric absorption features (chiefly in the red wing). The correction to the heliocentric velocity is --9.775\,\kms.}\label{fig-profiles-v1309}
\end{figure}

The line profiles of CO and SiO are shown in Fig.\,\ref{fig-profiles-v1309}. The Band\,7 lines, which were observed with a good S/N, have overall a Gaussian shape but their red wings are slightly stronger than the blue ones. Gaussian fits to the entire profiles indicate an average LSR velocity of --82$\pm$7\,\kms\ (heliocentric --92\,\kms), but the apparent maxima are slightly blue-shifted with respect to this value, i.e. are at about --100\,\kms\ (LSR). The main peaks of weaker unblended profiles of SO$_2$ lines are centered at similar velocities as CO and SiO. The FWHMs of the molecular lines are of 210 and 236\,\kms\ and the emission extends from --300 to 125\,\kms\ at their bases. The CO 3--2 line is slightly broader than SiO; CO 2--1 appears even broader still but at its modest S/N the exact extent and shape of this line is uncertain. 

In Fig.\,\ref{fig-profiles-v1309}, we compare the profiles of rotational lines observed with ALMA to these of electronic atomic transitions observed in the visual range with Xshooter at the Very Large Telescope. The X-shooter spectra were acquired between May and July 2016, i.e. a few months after the ALMA observations in Band\,7 and one year after those in Band\,6. Visual line profiles of V1309\,Sco in earlier epochs (2009--2012) were analyzed in \citet{kamiV1309} and the spectra observed in 2016 bear very similar characteristics. Many of the observed lines, like those of H$\alpha$, the \ion{Rb}{I} doublet, and [\ion{Ca}{I} $\lambda$6572 are centered at similar velocities as the rotational lines but are narrower and lack the bump in the red-shifted wing seen in molecular lines. The overlap of molecular lines and that of H$\alpha$ may indicate that some volume of molecular gas is partially collocated with the gas that in 2016 was still recombining after the 2008 eruption. 

The lines of the \ion{K}{I} doublet have a similar width as the molecular emission but are slightly more blueshifted. Emission of [\ion{O}{I}] $\lambda$6300 and that in the quadruple transition of H$_2$, $\varv$=1--0 $Q$(1), are centered at even more negative velocities but still fall in the velocity range occupied by CO and SiO emission. Origin of this shocked material is unclear but may be a signature of an ongoing dynamic interaction within the remnant or between the ejecta and pre-existing material \citep{kamiV1309}.

From this comparison, it appears that the molecular gas seen in rotational transitions has a different spatio-kinematical distribution than that traced by visual and infrared transitions. There exist some substructure of the remnant with different relative contributions from shocked, neutral, and molecular gases. In particular, the molecular gas has a higher column density in the redshifted part of the outflow compared to the blue one, in contrast to what is seen in some atomic lines and the H$_2$ line.

\begin{figure*}[!h]\sidecaption
\includegraphics[width=6cm]{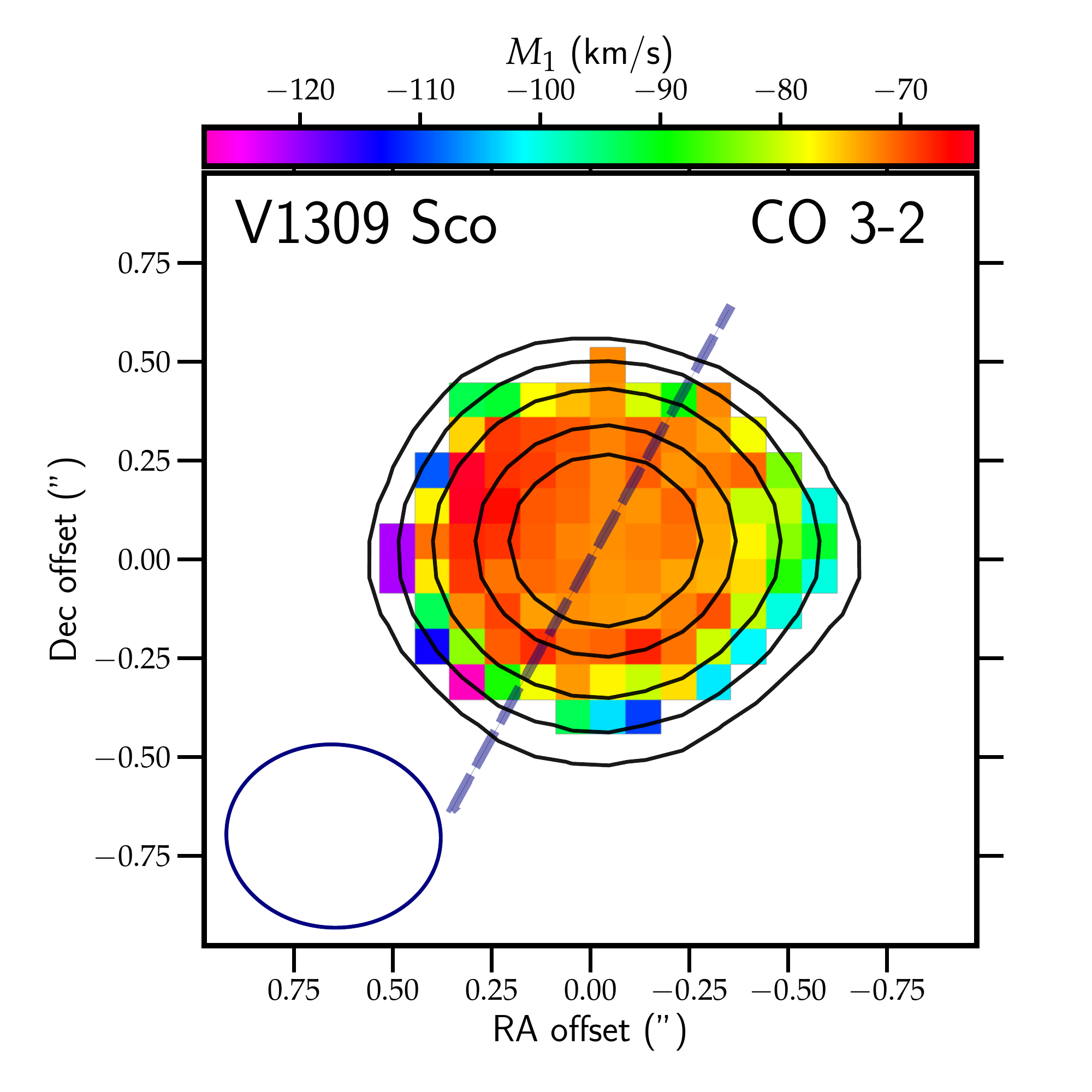}
\includegraphics[width=6cm]{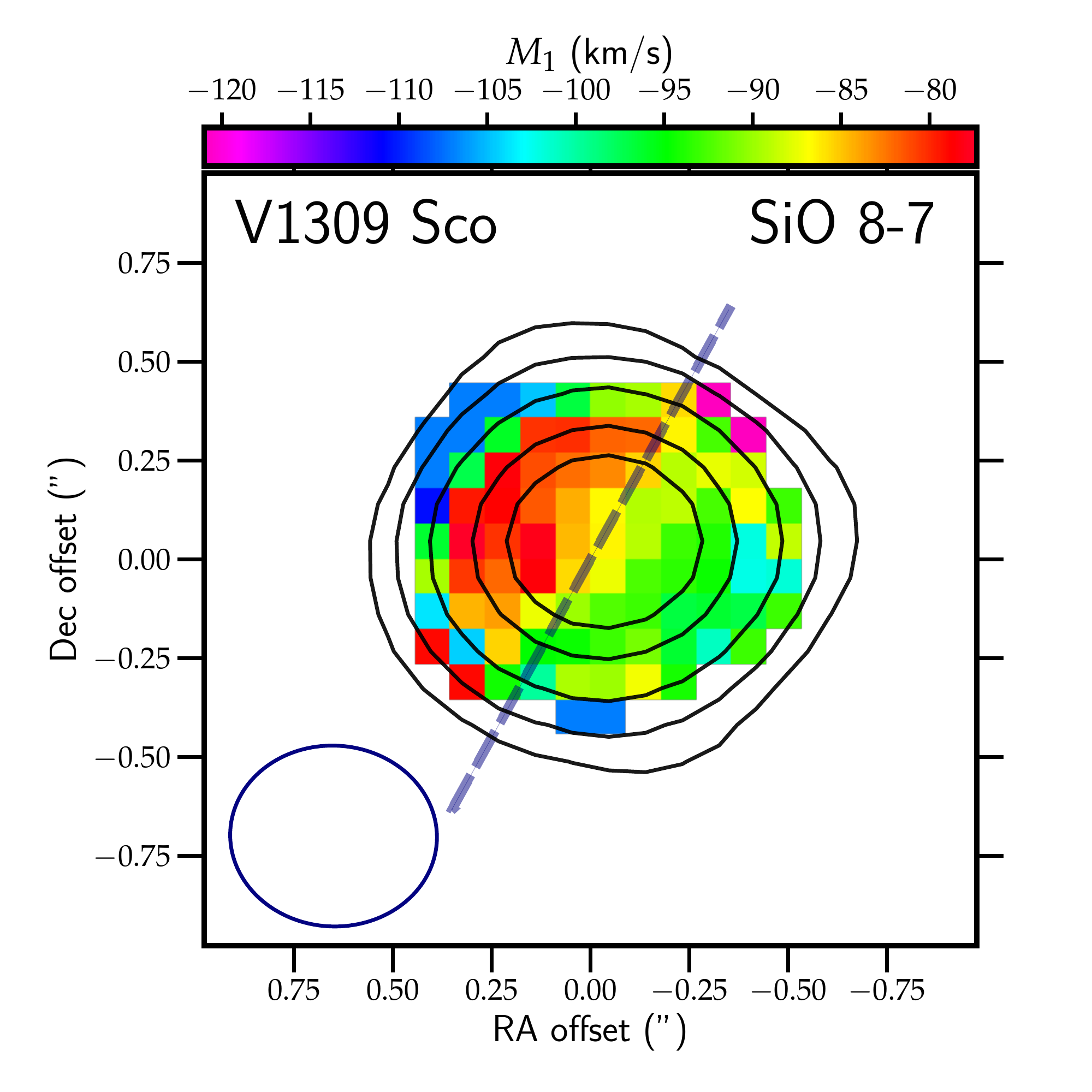}
\caption{First-moment maps and contours of the total intensity of CO 3--2 (left) and SiO 8--7 (right) in V1309\,Sco. The first-moment maps provide a view on the intensity-weighted mean velocity in the given pixel. The contours are drawn at 5, 10, 20, 40, 60, and 80\% of the respective emission peaks. Beam sizes are marked in the corners. The dashed line is drawn at a PA=151\degr\ along which the emission is slightly more extended.}\label{fig-mom1-v1309}
\end{figure*}

The spatial information on the kinematics is very limited owing to an insufficient angular resolution. The first-moment maps in Fig.\,\ref{fig-mom1-v1309}, especially that of SiO emission, indicate that the bulk of the emission in the north-eastern part is redshifted and south-western is blueshifted with respect to the average velocity. The spatio-kinematical structure of this source may be very similar to that of V4332\,Sgr observed with ALMA but the inclination angle of the bipolar structure may be even lower for V1309\,Sco. This notion is consistent with the constraints of the inclination of the progenitor binary orbit \citep[e.g. 84\degr\ for the orbit and 6\degr\ for an outflow orthogonal to the orbit;][]{PejchaBipolar}. However, the emission regions in V1309\,Sco are slightly more extended in the direction perpendicular to the velocity gradient (Fig.\,\ref{fig-mom1-v1309}), in contrast to what is seen in V4332\,Sgr. This may be related to the younger age of the V1309\,Sco's remnant.

Assuming that the observed molecular gas was dispersed in the 2008 eruption of V1309\,Sco, we derive its kinematic distance. The emission size with a FWHM of 164\,mas and spectral profile with a FWHM of $\sim$220\,\kms\ observed 7.5\,yr after the main eruption indicate a distance of 2.1\,kpc. This is only slightly smaller than 3.0$\pm$0.7\,kpc derived by \citet{v1309}. At $\sim$2.5\,kpc, the largest $e$-folding radius (FWHM/2 $\approx$ 85\,mas) of molecular emission corresponds to 213\,AU.

\subsection{Excitation model of V1309\,Sco}\label{sec-cassis-v1309}
We simulated the spectrum of V1309\,Sco in a similar fashion as in Sect.\,\ref{sec-cassis-v4332}. The resulting simulation is shown in Fig.\,\ref{fig-cassisModel-V1309}.
Except for CO and SiO, the spectra were simulated with two Gaussian profiles centered at LSR velocities of --95 and 53\,\kms. For SO$_2$, the component at negative velocities was found to be best reproduced by a FWHM of 160\,\kms\ and the positive weaker component by a FWHM of 90\,\kms. The numerous lines of SO$_2$ allowed us to constrain the excitation temperature of 78$^{+10}_{-8}$\,K for the negative component and 81$^{+42}_{-21}$\,K for the positive one, i.e. within the errors both components have equal temperatures. The same temperatures and line widths were then adopted to simulate the spectra of SO, AlO, and SiO. The resulting column densities are listed in Table\,\ref{tab-Ncol-v1309}. The two components appear to have different relative abundances of SO, AlO, and SO$_2$. The lack of emission from isotopologues containing rare isotopes of sulfur is consistent with solar isotopic abundances.

\begin{table}\caption{Column densities (in cm$^{-2}$) of species observed in V1309\,Sco.}\label{tab-Ncol-v1309}
\centering\small\begin{tabular}{ccc}
\hline
Molecule & \multicolumn{2}{c}{Component at}\\
         & --95\,\kms & 53\,\kms \\
\hline
SO$_2$& $4.60 \times 10^{16}$ & $1.20 \times 10^{16}$ \\
SO    & $1.85 \times 10^{16}$ & $4.60 \times 10^{15}$ \\
AlO   & $5.90 \times 10^{14}$ & $1.22 \times 10^{14}$ \\
AlOH  & $<2.5 \times 10^{15}$ & \\
\hline
 & \multicolumn{2}{c}{Single component}\\
\hline
CO\tablefootmark{a}   & \multicolumn{2}{c}{$2.40 \times 10^{18}$}\\
SiO\tablefootmark{b}  & \multicolumn{2}{c}{$2.60 \times 10^{15}$}\\
\hline
\end{tabular}\tablefoot{
\tablefootmark{a} Centered at an LSR velocity --72.5\,\kms.
\tablefootmark{b} Centered at an LSR velocity --82.0\,\kms.
 }
\end{table}

Observations of two transitions of CO allowed us to independently constrain the temperature of the CO-bearing gas. Although the Band\,6 and 7 observations were obtained about six months apart, we ignore any potential variability in the rotational lines over this time. We modeled the emission as arising from isothermal gas of a single Gaussian velocity profile. The LTE calculation yields an excitation temperature of 36$^{+26}_{-8}$\,K and a column density of 2.4($\pm$0.3)$\times$10$^{18}$\,cm. The derived temperature is approximately half that derived for SO$_2$. The column density indicates a total mass of H$_2$ and He (He at solar abundance) of $4\times 10^{-4}$\,M$_{\sun}$. We assumed CO-to-H$_2$ number ratio of $3 \times 10^{-4}$ and a distance of 2.5\,kpc. Because the emission may be saturated, the column density of CO and the total mass estimate should be considered as lower limits. This mass is much lower compared to that of the dust in the ejecta  of 10$^{-3}$\,\msun\ derived from SED analysis in \citep{v1309sed}, which at the standard dust-to-gas mass ratio of 100 would indicate a total ejecta mass of 10$^{-2}$\,\msun (see discussion in Sect.\,\ref{sec-masses}).

Although the $^{13}$CO $J$=1--0 transition was covered by the Band\,3 spectrum, its  sensitivity  is inadequate to put useful constraints on the $^{13}$CO abundance relative to $^{12}$CO.

\section{V838 Mon}\label{sec-V838}
\subsection{SMA observations}
We observed V838\,Mon with the SMA on three dates. First observations were taken on 15 Oct. 2016 with the 345 and 400 receivers and covered four spectral ranges: 296.9--303.1, 310.8--317.0, 332.8--339.0, and 346.8--353.0\,GHz. The bandpass was calibrated by observations of J1058+015. The calibration of gain phases was obtained through cyclic observations of J0725-009 and J0730-116. Only the former quasar was used for calibrations of gain amplitudes. Although Vesta was observed to set the absolute flux scale, its model fluxes turned out to be too inaccurate. We used instead fluxes of our bandpass calibrator interpolated to the observed dates and frequencies from measurements in the ALMA calibrator catalog\footnote{\url{https://almascience.eso.org/alma-data/calibrator-catalogue}}. We expect that this resulted in the absolute flux calibration better than 30\%.

On 18 and 28 November 2016, we used the 230 and 400 receivers of the SMA to cover the spectral ranges: 213.4--219.5, 227.4--233.6, 330.7--336.8, and 344.7--350.9\,GHz. The latter two ranges partially overlap with the observations on 15 Oct. The bandpass calibration was performed with observations of 3C273 and the absolute flux scale was calibrated with Callisto. With the aim to increase the astrometric accuracy of the observations, instead of the usual two gain calibrators strategy, we observed four nearby quasars, 0725-009, 0730-116, 0739+016, and 0607-085. They were all used for calibration of phases while amplitudes were calibrated on 0725-009 only. The overall flux calibration is expected to be better than 20\%.

All the SMA observations were obtained with the newly installed SWARM correlator which at the time had three functional quadrants per sideband. The performance of edge channels of each quadrant was not optimal, especially in November 2016. In consequence, the reduced spectra have narrow gaps between the different quadrants. The raw data were registered with a resolution of 140\,kHz but were rebinned to a lower resolution for processing. Here we present the spectra at a resolution of $\sim$30\,\kms.

The SMA observations were obtained with 7 to 8 antennas arranged in the extended configuration with baselines between 44 and 226\,m. At natural weighting of the V838\,Mon's visibilities, this configuration resulted in a beam FWHM of 4\farcs3$\times$1\farcs1 at 214\,GHz and 0\farcs9$\times$0\farcs6 at 353\,GHz. The corresponding representative rms noise levels per 30\,\kms\ bins are 7.1 and 27.1\,mJy per beam, respectively.

No continuum emission was convincingly detected in V838\,Mon. Fitting a straight line to spectral ranges free of obvious emission lines yielded emission at a level of 2.5$\pm$0.3\,mJy/beam in the 230 receiver data and 10$\pm$9\,mJy/beam in the combined data from receiver 400 (the uncertainties correspond to 1$\sigma$ errors). It is however unlikely that the emission at $\sim$1.3\,mm is a true continuum -- it is rather cumulative flux from molecular lines observed at a very low S/N. These derived fluxes should therefore be treated as upper limits on the continuum emission at 223 and 342\,GHz. These flux levels are negligibly low compared to the lines we analyze here (above 0.5\,Jy/beam) and the spectra were not corrected for the potential presence of continuum. All data were calibrated in MIR\footnote{\url{https://www.cfa.harvard.edu/~cqi/mircook.html}} and imaged in CASA.

\subsection{Identification of spectral features}
\begin{figure*}[!h]
\includegraphics[angle=270, width=0.99\textwidth]{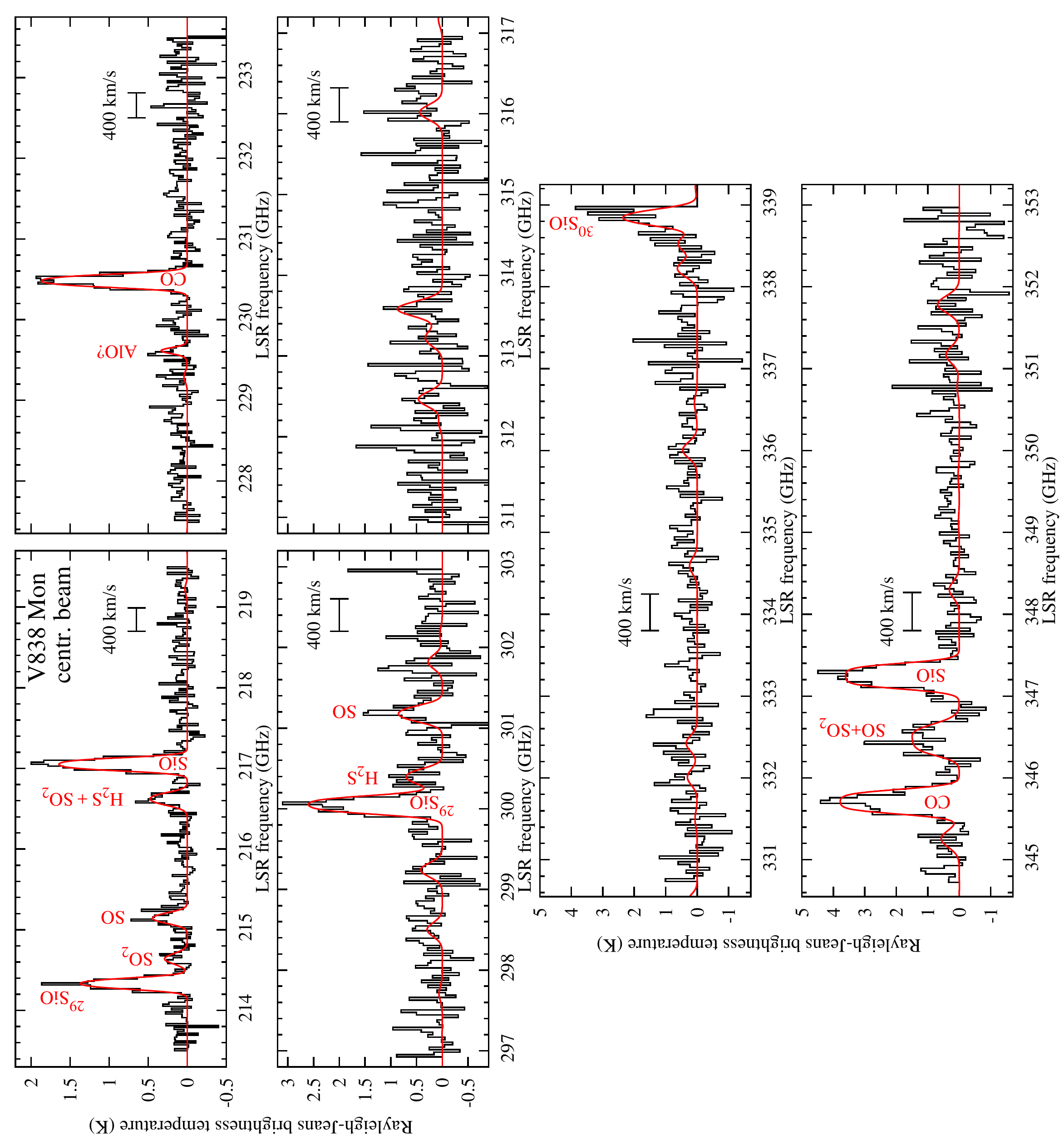}
\caption{The SMA spectra of V838\,Mon (black) and their LTE simulation (red). Obsered data were smoothed to a resolution of $\sim$30\,\kms.}\label{fig-cassisModel-v838}
\end{figure*}

Although the SMA spectra have lower sensitivity than the ALMA data presented in Sects.\,\ref{sec-V4332} and \ref{sec-v1309}, their coverage is much wider and more molecules could be observed in V838\,Mon. Over a dozen of transitions have been identified in the SMA spectra. They are indicated in Table\,\ref{tab-lines}. Earlier far-infrared observations on {\it Herschel} revealed the presence of thermal lines of H$_2$O, CO, and SiO in V838\,Mon \citep{exter}. While we do not cover any strong lines of water, identifying spectral features of CO and SiO in the SMA spectra was straightforward. The two sulfur oxides present in the other two red novae are also present in V838\,Mon but emission of SO$_2$ is very weak. There is no conclusive evidence for the presence of AlO but a very weak feature is present close to the position of the $N$=6--5 line near 229.8\,GHz. Moreover, we identified emission of the two rare isotopologues of SiO, $^{29}$SiO and $^{30}$SiO, and two emission features of H$_2$S.\footnote{Third prominent line of H$_2$S, 6,1,6--5,0,5, near 1846.77\,GHz was found in the PACS spectra of \citet{exter} (their Fig.\,2).} The SMA spectra are shown in Fig.\,\ref{fig-cassisModel-v838}.

\subsection{Location and kinematics of molecular gas}\label{sec-location-v838}
The source of emission is centered at J2000 coordinates RA=07$^h$04$^m$04\fs8200 ($\pm$6\,mas) and Dec=--03\degr50\arcmin50\farcs635 ($\pm$4\,mas) and is unresolved in our SMA observations. The location of the submm source is close to the coordinates of the SiO maser, RA=07$^h$04$^m$04\fs824 and Dec=$-03$\degr50\arcmin50\fs50 ($\pm$10\,mas), measured by \citet{claussen} with the Very Large Array (VLA) in 2005\footnote{We assume there is a typo in the RA minutes in the cited paper.}. The difference in positions is significantly larger than the uncertainties, possibly due to an error in the VLA position. A Very Long Baseline Interferometry observation of the SiO maser from 2017 agrees very well with the position we obtained with the SMA (G. Ortiz, priv. comm.).

\begin{figure}
\includegraphics[angle=270, width=0.99\columnwidth]{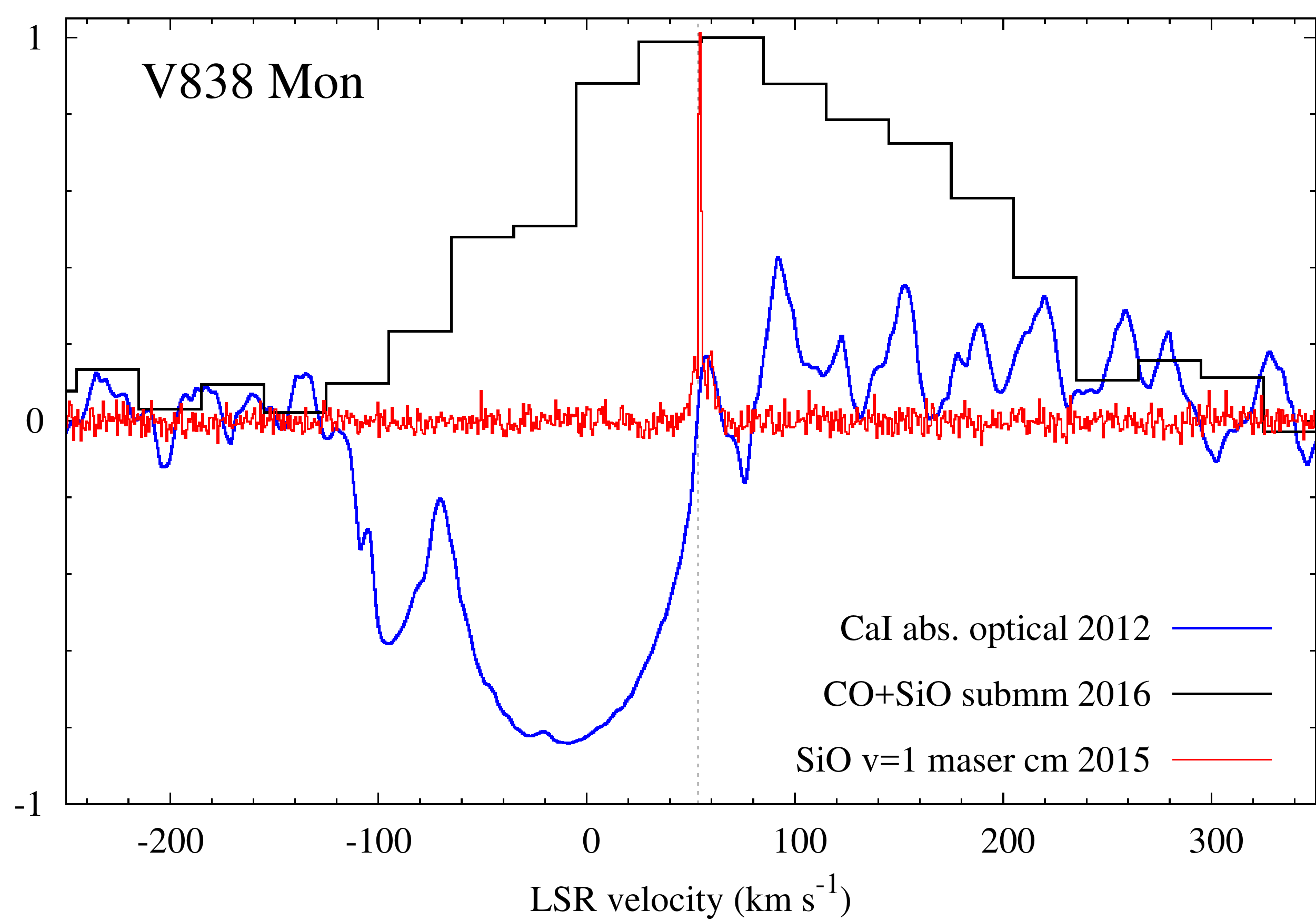}
\caption{A mean profile of molecular emission observed with the SMA (black) in V838\,Mon is comapred to the SiO $\varv$=1 $J$=2--1 spectrum and an optical absorption profile of \ion{Ca}{I} at 6573\,\AA. The profile of molecular emission is a mean of normalized unblended lines of CO and of $^{28,29}$SiO observed in 2016. The maser emission near 86\,GHz was observed with the IRAM 30\,m telescope in 2015. The maser defines the systemic velocity of the star which is marked with a vertical dashed line. The absorption profile of \ion{Ca}{I} was extracted from a spectrum obtained in 2012 with VLT/UVES; it was normalized and shifted by --1 . Its deep double absorption trough represents circumstellar absorption on the line of sight toward the star. A wavy pattern in the optical spectrum is caused by molecular absorption features, mainly by electronic bands of TiO. }\label{fig-maser}
\end{figure}

Although peaks of the strongest lines appear at 40--70\,\kms, fitting Gaussian profiles to overall profiles of CO and SiO gives central LSR velocities between 72--84\,\kms. Lines of SO are centered at an even higher velocity of $\sim$107$\pm$24\,\kms. The large scatter in measured velocities is caused by the modest S/N and asymmetries in the lines. Although the asymmetries could be partially caused by contamination from other unidentified species, the red wings are consistently stronger in all the prominent lines. The peak of an averaged SiO and CO profile at 77$\pm$4\,\kms, is red-shifted with respect to the radio SiO masers of V838\,Mon \citep{claussen,deguchi}. In Fig.\ref{fig-maser}, we compare the average SiO and CO profile to that of the SiO $\varv$=1 $J$=2--1 maser from IRAM-30\,m observations in 2015 (Appendix\,\ref{sec-past}). The maser emission arises within a few stellar radii from the photosphere of the stellar remnant of V838\,Mon and is likely associated with a post-outburst wind of the giant star. Its central velocity of 54$\pm$1\,\kms\ is assumed to be the systemic velocity of V838\,Mon itself \citep{deguchi}. The 23\,\kms\ mismatch between centroids of mm/submm emission and that of the maser can be understood if the mm/submm lines are optically thick and/or represent gas with an asymmetric distribution and complex velocity field. Optical observations of the V838\,Mon remnant also suggest asymmetries in the remnant.  Because late post-outburst visual spectra show no pure emission features, the visual studies are limited to absorption lines which probe only the line of sight toward the star exposing the innermost regions of the envelope (Fig.\ref{fig-maser}) or the supergiant pseudo-photosphere \citep{tylFeII}. From the mm/submm spectra, it appears  that the far side of the circumstellar envelope, not accessible in the visual studies, is richer in cool material than the near  side. Alternatively, some self-absorption within the molecular profiles may produce the observed asymmetry.

The typical FWHM of the molecular lines is 230$\pm$10\,\kms, i.e. comparable to or somewhat smaller than widths of lines observed during and after the 2002 outburst at optical and infrared wavelengths \citep{kolka,kamiKeck}.  Adopting the distance to the star of 6.1\,kpc \citep{sparks}, taking the typical expansion velocity as FWHM/2=115\,\kms, and assuming that tangential motions are similar to these traced in radial velocities, we calculate that $\sim$15\,yr after the eruption\footnote{Although the object was discovered in January and brightest in April 2002, the star was very likely disturbed and experiencing mass loss already in December 2001 \citep{tylEvolV838}.} the circumstellar remnant had an extent of about 117\,mas (715\,AU), in the sense of a Gaussian FWHM. This figure is slightly larger than $\lesssim$70\,mas derived from mid-infrared interferometric measurements of continuum in 2011 \citep{olivier} but consistent with continuous expansion of the structure and considering large uncertainties in both estimates. We note, however, that full line widths of SiO and CO of 405\,\kms\ suggest expansion velocities of 203\,\kms\ which would indicate an emission region of a full size of 206\,mas.

\subsection{Excitation conditions and mass constraints}
Using CASSIS procedures similar to these used for the two other objects (Sects.\,\ref{sec-cassis-v4332} and \ref{sec-cassis-v1309}), we constrained the column densities and excitation conditions for all the detected molecules in V838\,Mon. The temperatures derived independently for each species range from 70--150\,K for SO, CO, SiO to $>$210\,K for SO$_2$. Our models and the adopted source size imply a very high column density of 1.1$\times$10$^{20}$\,cm$^{-2}$ for CO. The column density of SiO, 1.8$\times$10$^{17}$\,cm$^{-2}$, is also relatively high. Our correction for line saturation at such high concentrations is likely inadequate (see below) and the column densities should be treated as upper limits only. Among the sulfur oxides, SO$_2$ with $N_{\rm SO2}\gtrsim 1.0 \times 10^{18}$\,cm$^{-2}$ is more abundant than SO with $N_{\rm SO}$=2.5$\times$10$^{17}$\,cm$^{-2}$. The third S-bearing species, H$_2$S, has intermediate abundance with a column density of 7.2$\times$10$^{17}$\,cm$^{-2}$. Assuming an excitation temperature of 150\,K for AlO, we derive an upper limit on its column density of $\sim$4.0$\times$10$^{15}$\,cm$^{-2}$. The full simulation of the spectrum is shown in Fig.\,\ref{fig-cassisModel-v838}.

We also attempted to constrain the abundance of the isotopologues $^{29}$SiO and $^{30}$SiO relative to the main species, $^{28}$SiO. We included each of these two rare species in the model of SiO with an abundance ratio as a free parameter. The models yield a relative SiO-to-$^{29}$SiO abundance of 1.6$\pm$0.4 (1$\sigma$) and that of SiO-to-$^{30}$SiO of 4.3$\pm$2.1. In solar composition and with no isotopic fractionation, these ratios are of 19.7 and 29.9, respectively. Although an isotopic anomaly cannot be excluded for V838\,Mon, the low derived ratios may rather indicate a very high saturation in the SiO lines, which our model did not adequately correct for, or a maser component. Maser emission is certainly present in lines at excited vibrational levels observed at radio wavelengths.

A lower limit on the mass of the cool material traced in CO emission is 0.1\,\msun. For the calculation, we assumed an CO abundance of $f_{CO}$=3$\times$10$^{-4}$ relative to H$_2$, took into account the presence of He at solar elemental abundance, and assumed a source size of 117\,mas. The highest uncertainty in this estimate comes from the value of $f_{CO}$ (see below). As pointed out in Sect.\,\ref{sec-location-v838}, however, also the source size is rather uncertain.  Adopting a source size of 206\,mas whose solid angle is equivalent to a Gaussian with a FWHM of 172\,mas, we obtain a CO column density of 2.8$\times$10$^{19}$\,cm$^{-2}$ and a corresponding total mass of 0.05\,\msun. This is our more restrictive lower limit on the mass dispersed in the eruption of V838\,Mon.

\section{Discussion}\label{sec-discussion}
\subsection{The origin of the molecular material}
The extremely large widths of the lines observed in the remnants of the three objects indicate that the molecular material was ejected during their recent explosions as red novae. Observations during the eruptions indicate that the ejected material was ionized and had temperatures of about 10$^4$\,K. After the eruption the stellar remnant of red novae and their circumstellar environments cool down to very low temperatures. Molecular absorption bands appear very early on in optical spectra after the outburst, typically within months \citep[e.g.][]{mason}. It is therefore reasonable to interpret the molecular gas observed with ALMA and SMA as the component of the ejecta that cooled down to temperatures of 30--200\,K and recombined to molecular phase. How early after the eruption can we observe mm/submm emission in red novae? In Appendix\,\ref{sec-past}, we review the past attempts to detect rotational lines in the three red novae. Unfortunately, most of them were unsuccessful owing to too low sensitivity or inadequate observation techniques so we are unable to reconstruct the flux changes in these lines. 

\subsection{Masses dispersed in the merger events}\label{sec-masses}
Earlier literature estimates of the mass lost during a red nova event have been performed primarily for V838\,Mon. \citet{tylEvolV838} analyzed all available observations from the outburst and the immediate post-outburst phase of V838\,Mon and -- based on dynamical considerations -- put a very rough upper limit on the ejected mass of 0.6\,\msun, emphasizing that the ejecta is most likely much less massive than this. Far-infrared molecular lines and continuum emission observed by {\it Herschel} in 2011 and analyzed in \citet{exter} were modeled as arising from 0.2\,\msun\ of cool material. Our estimates, derived using similar methods, imply the thus-far lowest value on the dispersed mass in V838\,Mon of about 0.1\,\msun. Concerning other red novae, as mentioned in Sect.\,\ref{sec-cassis-v1309} the SED analysis of the V1309\,Sco remnant implies a mass that is two orders of magnitude larger than our ALMA constraint of 10$^{-4}$\,\msun. The discrepancy may be related to an unusual dust to gas mass ratio in V1309\,Sco.

The minimum masses we derive are accurate to within a factor of about $\sim$2, at best, owing to uncertain conversion factor between the number density of CO and H$_2$, $f_{CO}$. For classical oxygen-rich envelopes, e.g. these of red supergiants and M-type Miras, the fractional abundance of CO reported in the literature varies in a range between 1$\times$10$^{-4}$ and 6$\times$10$^{-4}$ \citep[e.g.][]{KnappMorris1985}. It is very challenging to constrain this ratio through observations for individual objects and there may be real variations between different stars owing to their chemical evolution. With no information on this ratio in the three red novae, we adopt a typical value of $f_{CO}$=3$\times$10$^{-4}$. In solar composition, complete depletion of carbon to CO molecules would result in $f_{CO}$=5.8$\times$10$^{-4}$. Our mass estimates would be nearly twice lower at this value of $f_{CO}$. However, non-equilibrium shock-driven chemistry that presumably took place in these explosive objects is known to produce C-bearing species on the cost of CO in O-rich media \citep[e.g.][]{gobrecht} and therefore $f_{CO}<5.8 \times 10^{-4}$ even if the elemental composition is solar.

The ejecta masses derived here, of 0.05, 0.01, 10$^{-4}$\,\msun\ for V838\,Mon, V4332\,Sgr, and V1309\,Sco, respectively, range over two orders of magnitude. The differences may be partially owing to the difference in the ages of the remnants. The gas dispersed in the youngest remnant of V1309\,Sco, is still effectively cooling down in high-excitation lines at optical wavelengths, including recombination lines \citep{kamiV1309}, and a large portion of the gas is likely still in atomic (ionic) form. Future monitoring of the mm/submm molecular lines in V1309\,Sco can help us verifying this hypothesis -- we predict that there should be an increasing mass of molecular material as the remnant of V1309\,Sco cools down. In the two older red novae, some atomic material is certainly also present as it is clearly seen in emission lines in V4332\,Sgr and in deep absorption lines in V838\,Mon. The masses of these atomic components are currently unknown.

Another reason for the large differences in ejecta masses in the three red novae may be related to the progenitor masses. Based on the remarkable similarity in post-outburst characteristics of V4332\,Sgr and V1309\,Sco \citep{kamiV1309}, we assume here that V4332\,Sgr was an identical  binary as that destroyed in the red nova event of V1309\,Sco, i.e. a system with a mass of approximately 1.54+0.16\,\msun\ \citep{stepien}. In case of V838\,Mon, the merged stars had masses of 8.0$\pm$3 and 0.3$\pm$0.2\,\msun\ \citep{tylendaProgenitor,TS06}. V4332\,Sgr was 22--23\,yr and V838\,Mon was 14\,yr after their eruptions when the ALMA observations were taken so both have had more time than V1309\,Sco to cool down to temperatures at which a considerable fraction of the dispersed material has a molecular form. The ejected mass constitutes then 1\% and 0.6\% of the mass of the progenitor system in V838\,Mon and V4332\,Sgr, respectively. 

Theoretical studies of head-on collisions between non-rotating stars indicate that the physical parameters that most strongly influence the mass of ejected material are the internal density structure of the stars and their relative mass ratio, $q$ (defined as the mass ratio of the lower-mass companion to the higher-mass star) \citep{lombardi}. In a simplified formula of \citet{GlebbeekPols}, the ejected mass fraction (relative to the mass of the binary) is $\varphi$=0.3$q/(1+q)^2$. For V838\,Mon ($q\!\approx$0.038) this prediction gives $\varphi$=1\% and for V4332\,Sgr with $q$=0.58 the formula gives $\varphi$=6\%. Both estimates are relatively close to the limits derived from our observations, considering the uncertain mass of the V4332\,Sgr's progenitor. This agreement is somewhat surprising, as the red-nova mergers were almost certainly not head-on central collisions and the binary of V4332\,Sgr had probably evolved beyond the main sequence. The close match may be coincidental and future studies of other Galactic red novae can define the observational relation for $\varphi$. Other theoretical studies focused on mergers in systems similar to V1309\,Sco predict masses of unbound material of 0.038--0.086\,\msun\ \citep{nandez} and $>$0.05\,\msun\ \citep{PejchaBipolar}, consistent with our lower mass limit for V1309\,Sco. Overall, the ejected mass in a stellar merger -- as constrained by observations -- is very small, of a few percent of the total mass of the catastrophic binary.

The ejecta masses derived here of 10$^{-4}$--0.1\,\msun, although constitute lower limits only, are typically already larger than ejecta masses of classical novae which are  in the range 1--10$\times$10$^{-5}$\,\msun\ \citep{classicalNovae}. Also, no cool molecular material has been ever identified in classical novae years after their eruption \citep{kamiNat}. The high mass of cool molecular gas can be used as a distinguishing feature of red novae among other eruptive stars. 

\subsection{The structure and kinematics of the ejecta}\label{sec-discussion-kine}
Although the mass ejected during (or just prior to) a merger is low, it is expected to carry away a considerable amount of the angular momentum and orbital energy of the decaying binary \citep{nandez}. Unfortunately, we are not able to measure directly the angular momentum stored in the molecular remnants with the current observations. Predicting the details of how merging systems dispose of angular momentum is very challenging. In particular, if the angular momentum is taken away with the merger-burst ejecta, it is not clear what geometric configuration it would take. Many studies predict that a large fraction of the material is lost in the plane of the orbit. For instance, models of \citet{nandez}, \citet{Pejcha16A}, and \citet{morgan2018} predict strong mass loss through the outer Lagrange point, $L_2$, before the actual merger \citep[see also][]{iaconi,PejchaBipolar}. In simulations, this material forms a spiral-like circumstellar remnant which settles in a form of a ring after the merger is complete \citep{nandez}. Other studies, on the other hand, predict the formation of an accretion disk (or torus) which launches a pair of jets during the merger event \citep[e.g.][]{akashiJets,ILOTclass}. Although this jets scenario predicts mass loss in the orbital plane too, it is the jets that transfer away the excessive angular momentum from the remnant and form bipolar ejecta \citep{jetsILOTs}. Simulations of \citet{PejchaBipolar} show that an interaction of material accumulated in the orbital plane before the merger and that lost during the merger as spherically symmetric ejecta can result in a configuration spatially dominated by a wide-angle bipolar outflow. This latter scenario seems to be well supported by the observations of V4332\,Sgr and V1309\,Sco presented here.

The ALMA maps of V4332\,Sgr indicate that the remnants, or at least their molecular component, have axial symmetry consistent with a wide-angle bipolar outflow. The presence of jets is not excluded by the data. Whether any highly collimated fast-moving structures were formed, can be verified with future observations at higher angular resolutions and when the ejecta expands to larger dimensions. \citet{jetsILOTs} predict a linear velocity field in the form $\varv \propto r$, which is consistent with what we observe for the bulk of the molecular gas in V4332\,Sgr. We do not observe directly with ALMA any signatures of flattened structures such as rings or disks but the biconical symmetry we observe may indirectly point towards a presence of such structures. In planetary nebulae, post-AGB objects (pre-PNe), and $\eta$\,Car such bipolar structures are associated with a form of a torus or a disk which could have played a role in creating the bipolar structure. Similar scenario was proposed for V1309\,Sco in \citet{PejchaBipolar}. Additionally, the existence of disks or disk-like structures in V1309\,Sco and V4332\,Sgr has been advocated based on optical and infrared observations of these remnants \citep[e.g.][]{kamiV4332_2010,kamiV1309}. 

At the estimated masses here, the kinetic energies of the molecular ejecta are of the order of 10$^{46}$\,erg for V4332\,Sgr and V838\,Mon and of 10$^{44}$\,erg for V1309\,Sco. For V1309\,Sco, the energy is comparable to that radiated during its outburst \citep{v1309} but still much lower than the orbital energy ($\sim$10$^{47}$\,ergs). In the case of the older remnant of V838\,Mon, the energy stored in the molecular outflow is a large fraction of its expected orbital energy (however, the configuration of the progenitor binary is not well constrained). This leaves little doubt that the molecular gas was ejected in the outburst of each of these objects. The energies derived here are also comparable to these characterizing outflows of post-AGB objects -- e.g. \citet{Buja2001} found 10$^{44-47}$\,erg for a sample of 28 such objects. This reinforces the similarity of red novae to these evolved objects emphasized in Sect.\,\ref{intro}.

In the case of V838\,Mon whose systemic radial velocity is well known owing to the presence of SiO masers, it is clear that the centers of submm emission lines are considerably red shifted and could be misleading if used alone to determine the stellar systemic velocity. This is a consequence of the asymmetric geometry of the outflow with respect to the line of sight. The same is likely the case for the other two red novae studied here whose emission lines (optical and submm/mm) have central velocities that do not agree with the general Galactic velocity field (unless they would be halo objects) \citep{kimeswengerASP,kamiV1309,tyl2015}. The true systemic velocities of the two objects remain to be better constrained.

\begin{figure*}[!ht]
\includegraphics[angle=270, width=0.5\textwidth]{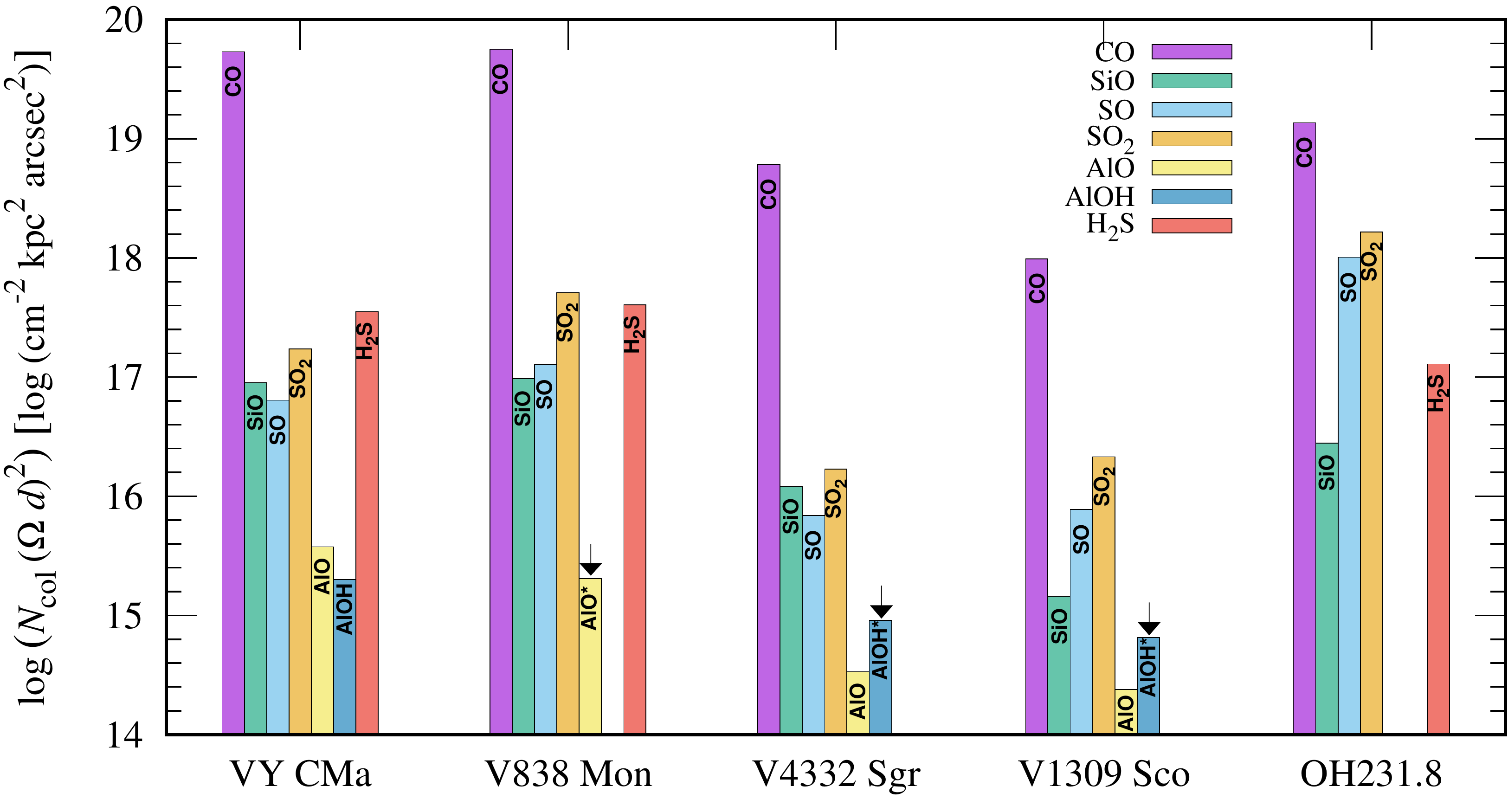}
\includegraphics[angle=270, width=0.5\textwidth]{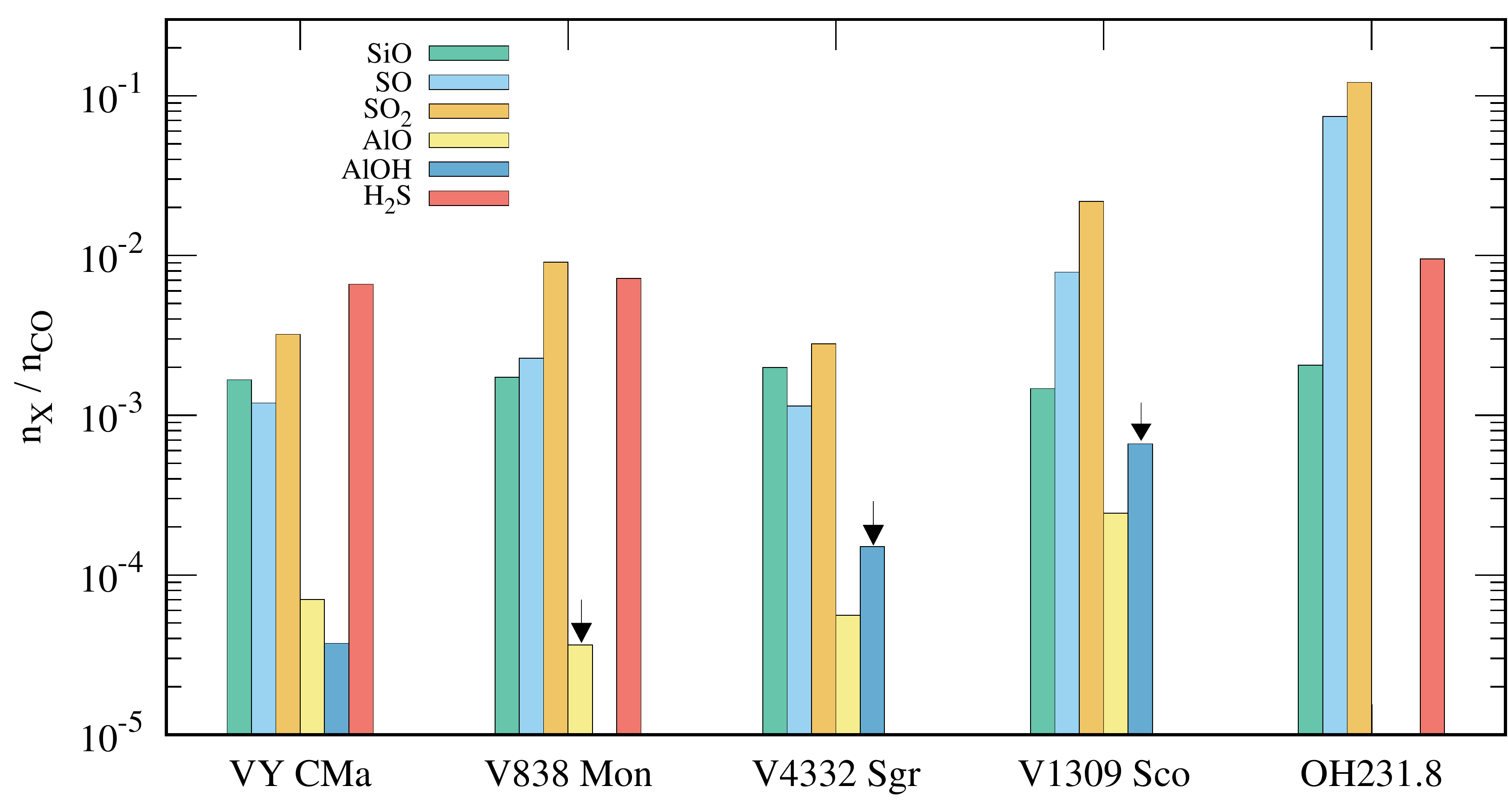}
\caption{{\bf Left:} Source-averaged numbers of molecules in the three red novea, VY\,CMa, and OH231.8. Upper limits are indicated with arrows and astericses in the labels. {\bf Right:} The same histogram but with molecule numbers normalized to that of CO in each object.}\label{fig-bars}
\end{figure*}

\subsection{Astro-chemistry of red novae}
In Fig.\,\ref{fig-bars}, we compare the source-averaged abundances of all the observed species in the three red nova remnants to these in the envelope of the red supergiant VY\,CMa and in the outflow of the post-AGB star OH231.8 (Rotten Egg Nebula). We show numbers of particles derived in this study for the red novae; in case of VY\,CMa, data were taken from \citet[][SiO and CO]{matsuura} and \citet[][all other species]{kami_surv}; and data of OH231.8 were taken from S{\'a}nchez Contreras et al.\footnote{\url{http://herschel.esac.esa.int/TheUniverseExploredByHerschel/posters/A88_SanchezContrerasC.pdf}}. The relative abundances are within uncertainties very similar in all three red novae. For instance, SO$_2$ is consistently three times more abundant than SO. 

The abundance patterns observed in the red novae are particularly close to these of VY\,CMa, a well-studied red supergiant with an O-rich and spatially-resolved ($\sim$10\arcsec) envelope. The envelope has a very complex morphology and displays multiple types of substructures (arcs, jets, filaments, knots, etc.) which suggest that different parts of the outflow interact with each other \citep{monnier,smithVY}. The chemistry of the gas is very likely also affected by these interactions. The close similarity of the derived abundances in VY\,CMa and red novae may suggest that the molecular remnants of red novae are equally complex and dynamic as the spatially resolved envelope of VY\,CMa. In the case of V838\,Mon, the chemical resemblance to VY\,CMa is very close even in terms of absolute numbers. The mass of the circumstellar molecular envelope of VY\,CMa is estimated as 0.15\,\msun\ \citep{muller}. The similarity between VY\,CMa and V838\,Mon is strengthened by the presence of SiO masers in both sources. With very comparable SiO abundances among all the red novae, why does only V838\,Mon show this maser emission? It is unclear what  in the immediate vicinity of the photospheres of V4332\,Sgr and V1309\,Sco stops them from producing a maser as that seen in V838\,Mon.  

The three red novae were the first astronomical objects towards which AlO had been detected through electronic bands at optical and near-IR wavelengths \citep{evansAlO,banerjeeAlO,kamiV1309}. The features of AlO have been considered as exotic and  interpreted as a signature of an AlO enhancement in these objects \citep{kamiV1309}. In comparison to VY\,CMa (Fig.\,\ref{fig-bars}), however, the relative abundances of AlO in red novae, as constrained here by emission in pure rotational transitions, do not appear that extraordinary at all. The source-averaged CO/AlO abundances of red novae of $\lesssim$3$\times$10$^{-4}$ are similar or below these in the superginat VY\,CMa \citep{kami_alo,kami_surv} and in the classical O-rich AGB star, $o$\,Ceti \citep{kamiAlOmira}. At solar composition, this leaves a large fraction ($\gtrsim$90\%) of elemental Al to exist in other forms, i.e. ionic, atomic, molecular and solid \citep[cf.][]{baner15}.  

\begin{figure}[!ht]
\includegraphics[angle=270, width=0.99\columnwidth]{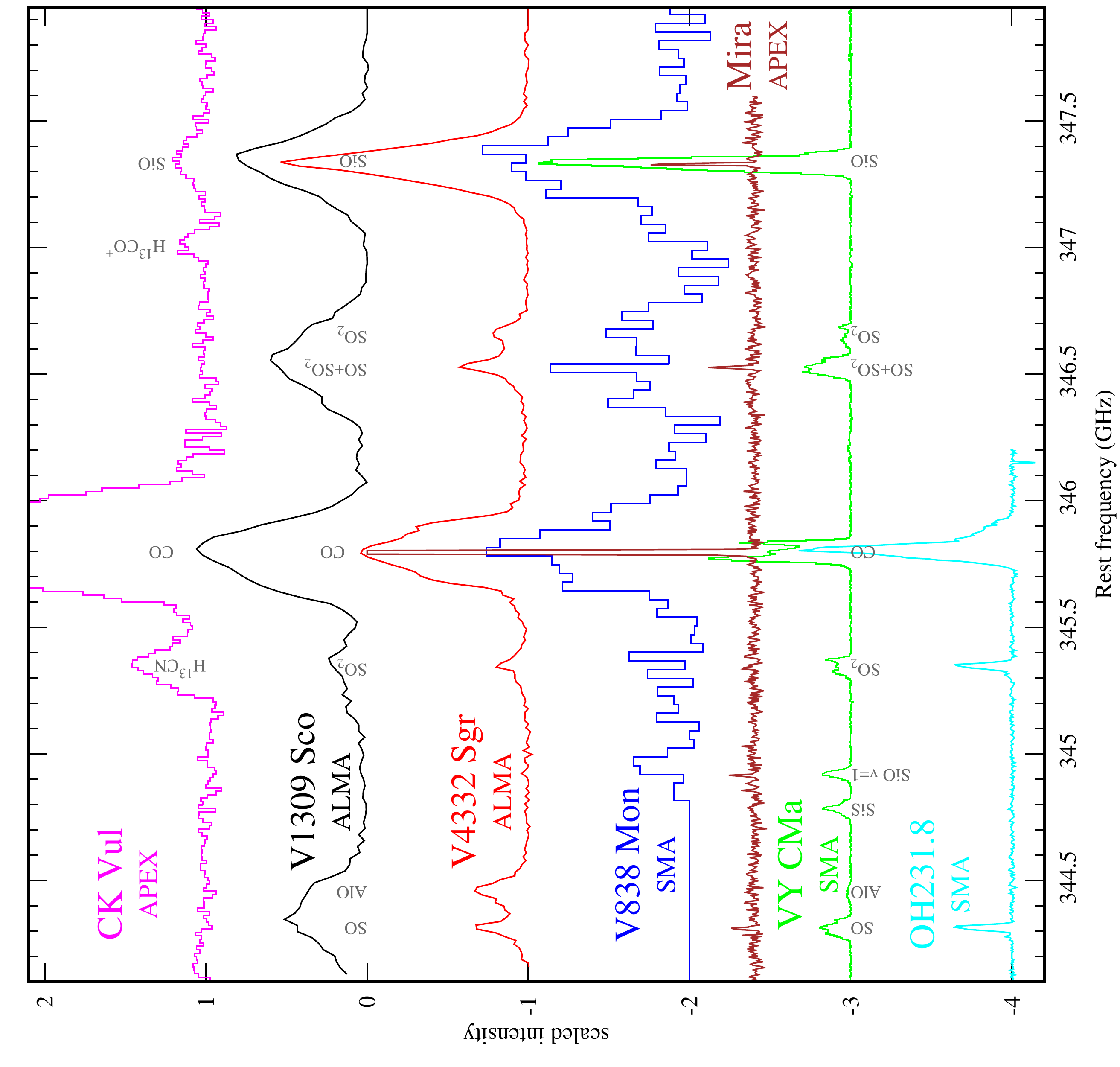}
\caption{Spectra of red novae are compared to these of classical circumstellar O-rich envelopes: a pulsating AGB stars, Mira ($o$\,Ceti, spectrum is from APEX ESO archive); a red supergiant, VY\,CMa \citep{kami_surv}; and a post-AGB object, OH231.8 (Rotten Egg Nebula) \citep{sabin}. The spectrum of CK\,Vul is from \citet{kami_ckvuldish}. The CO lines of Mira and CK\,Vul were trimmed for clarity.} \label{fig-allSpec}
\end{figure}

In Fig.\,\ref{fig-allSpec}, we compare sample spectra of red-novae remnants to spectra of the more classical oxygen-rich stellar sources of submm emission, including that of the red supergiant VY\,CMa, the AGB star $o$\,Ceti and the post-AGB object OH231.8. A spectrum of the carbon-enhanced remnant of CK\,Vul is also included in the figure. Except for CK\,Vul, the inventories of observable transitions are very similar in all objects and reflect a very similar chemical composition. The differences in the spectra are chiefly due to different excitation conditions (and, to a lesser degree, to different observation techniques). However, there are not many other objects displaying molecular emission with lines as broad as these in red novae. One prominent exception is the remnant of SN\,1987A which has a similar molecular inventory as these of the O-rich objects in \citep{ALMAsnr1987a} but has lines with full widths of over 2000\,\kms. 

Another object with broad molecular lines is CK\,Vul which is considered to be the oldest known object similar to red novae \citep{kato,tyl-blg360,kamiNat}. Compared to other red novae, however, the mm/submm spectrum of CK\,Vul is richer in C-bearing species and species containing rare isotopes of the CNO elements. Also, the excitation temperature of this remnant is only about 12\,K, lower than in the three objects studied here. Although the current study shows that CK\,Vul shares a lot of its observational characteristics with these more recent transients, it is not well understood why CK\,Vul displays such a distinctive chemical composition \citep[for more discussion, see][]{kami_ckvuldish}. We note however, that it is the only merger remnant observed hundreds of years after the eruption. Moreover, mergers can happen in different binary configurations, evolutionary stages, and between stars of different masses leading to a broad range of merger products, including such with extraordinary chemical composition.

\paragraph{Final words:} As mm/submm-wave sources, red-nova remnants constitute a new class of circumstellar environments that produce broad molecular emission lines with  full widths of 400\,\kms, i.e. higher than in AGB and post-AGB objects. The observations presented in this paper constitute only the first successful submm inteferometric observations of red novae. Future observations at higher sensitivities, angular resolutions, and broader frequency coverages will allow more accurate quantitative characteristics of the red-nova remnants. Other Galactic red novae can be observed at these wavelengths, including the least-studied transient OGLE-2002-BLG-360, for which no spectral features have ever been observed. Studies of these objects allow us to learn about stellar mergers, the most extreme case of binary interaction whose softer cases are very likely responsible for bipolar structures in post-AGB stars and planetary nebulae. 

\begin{acknowledgements}
We thank the anonymous referee for helping us improving the presentation in the paper. We thank L. Sabin and Q. Zhang for providing us with the SMA spectrum of OH231.8. We also thank L. Matr{\' a} for  fixing a problem with visibility weights in SMA data exported to CASA. We thank M. MacLeod for fruitful discussions on merging and common-envelope systems. We also appreciate our communications with N. Soker and A. Kashi on the role of jets in mergers. This paper makes use of the following ALMA data: ADS/JAO.ALMA 2015.1.00048.S and 2013.1.00788.S. ALMA is a partnership of ESO (representing its member states), NSF (USA) and NINS (Japan), together with NRC (Canada) and NSC and ASIAA (Taiwan) and KASI (Republic of Korea), in cooperation with the Republic of Chile. The Joint ALMA Observatory is operated by ESO, AUI/NRAO and NAOJ. The National Radio Astronomy Observatory is a facility of the National Science Foundation operated under cooperative agreement by Associated Universities, Inc. We used in this study data from the Submillimeter Array which is a joint project between the Smithsonian Astrophysical Observatory and the Academia Sinica Institute of Astronomy and Astrophysics and is funded by the Smithsonian Institution and the Academia Sinica. The study is also based on observations collected at the European Organisation for Astronomical Research in the Southern Hemisphere under ESO programmes 097.D-0092(A), 088.D-0112(B), and 085.F-9319(A). Part of the observations were carried out under project numbers 009-06, 054-15, and 058-11 with the IRAM 30m telescope. IRAM is supported by INSU/CNRS (France), MPG (Germany) and IGN (Spain). This research made use of NumPy \citep{numpy} and matplotlib, a Python library for publication quality graphics \citep{matplotlib}, and on the CASSIS software and JPL and CDMS spectroscopic databases. CASSIS has been developed by IRAP-UPS/CNRS.
\end{acknowledgements}.

\begin{appendix}
\section{Past observations of molecular rotational transitions in red novae}\label{sec-past}
\paragraph{V4332\,Sgr}
Observations of V4332\,Sgr in 2009 with the James Clerk Maxwell Telescope (JCMT) did not lead to detection although the spectra were centered on the CO(3--2) line. At the sensitivity of 3$\sigma \approx$613\,mJy per 0.42\,\kms\ and a beam size of 15\arcsec\ \citep{kamiV4332_2010} the emission observed with ALMA in 2016/2017 would not be detected. On 17 August 2011 a short-duration spectrum centered in CO 2--1 was taken with the IRAM 30\,m telescope (a beam of 11\arcsec) but the line was not detected at 3$\sigma$=650\,mJy per 20\,\kms. A search for molecular emission near the SiO 8--7 and CO 3--2 lines was also performed on 3--4 July 2012 with APEX telescope and FLASH-345 receiver. This receiver produces very flat baselines allowing for searches of broad lines. At a beam of 17\arcsec, a sensitivity of 3$\sigma$=291\,mJy per 20\,\kms\ was reached. It, again, was insufficient to detect even the strongest lines seen with ALMA.

\paragraph{V1309\,Sco}
IRAM 30\,m observations of V1309\,Sco in the CO 2--1 line on 22 Aug. 2011 resulted in a non-detection at 3$\sigma$=380\,mJy per 20\,\kms. Later, on 4 July and 30 Aug 2012, we used APEX and FLASH-345 to search for molecular lines in V1309\,Sco. A sensitivity of 3$\sigma$=173\,mJy per 20\,\kms\ was insufficient to detect even the strongest lines of CO and SiO seen by ALMA four years later.

\paragraph{V838\,Mon}
V838\,Mon has a rich history of failed attempts to detect mm/submm emission which, unfortunately, is not fully documented in the literature. \citet{kami_CO} and \citet{kami_coecho} summarized the first searches for mm/submm molecular emission towards V838\,Mon. These observations with single-dish antennas, including IRAM 30\,m, JCMT, and the Nobeyama 45\,m dish, did not reveal any rotational lines that could be ascribed to the immediate vicinity of the eruptive star but they revealed extended emission of cold molecular cloud associated with the light echo. We note also that the receivers and observational techniques used in these first observations were not always well suited to detect broad lines such as these seen by ALMA and SMA in red novae. V838\,Mon was later observed at APEX with APEX-2 receiver. A spectrum centered on the CO 3--2 line and acquired in 5--11 April 2010 did not result in any obvious detection at 3$\sigma$=384\,mJy per 20\,\kms. This sensitivity was close to that required to detect the CO 3--2 line observed with the SMA in 2016. An extra limiting factor in these observations was however a poor baseline performance of the APEX-2 receiver which did not allow us to distinguish broad lines from baseline ripples. 

Also, on many occasions, V838\,Mon was observed in the spectral regions near the classical SiO masers (i.e. near 43 and 86\,GHz) covering, among others, low-$J$ transitions of SiO at the ground vibrational level. One such observation was obtained with the IRAM 30\,m on 11 Aug 2015, relatively close in time to the SMA observations. It covered two spectral ranges, 83.1--91.0 and 98.9--106.6\,GHz, and a 1\,h-long integration resulted in a deep spectrum of the SiO $\varv$=1 $J$=2-1 maser (Fig.\,\ref{fig-maser}). By applying our CASSIS model based on SMA data to these archival spectra (correcting for the beam dilution) we found that these observations still lacked the sensitivity necessary to detect thermal emission of SiO and of other species corresponding to features seen in the SMA spectra. In other words, our model is consistent with the nondetections in the IRAM 2015 spectra, and other similar (less sensitive) observations in the past. 

In 2009, there were unsuccessful attempts to detect CO 1--0 from V838\,Mon with the CARMA array \citep{rottlerAAS}. Prior to our SMA observations, these were the only millimeter-wave interferometric observations of this source. Unfortunately, the sensitivity levels reached with CARMA was not reported. 

The first observational signatures of the molecular gas that we observed with the SMA were probably found in the FIR spectra from {\it Herschel} in 2011. \citet{exter} derived a gas temperature of 400$\pm$50\,K which is higher than what we derived for CO ($\sim$116\,K). Although effective cooling in the five years between the observations cannot be entirely excluded, the difference likely arises in systematic errors of both excitation studies. Our results are biased toward lower temperatures -- as we covered only the lowest rotational transitions of CO -- while the FIR spectra could overestimate the temperature by not accounting for these lower transitions and additionally are affected by the presence of the very extended cold gas component associated with the echoing cloud. For an assumed source size of 420\,AU (70\,mas at 6.1\,kpc), the {\it Herschel} CO emission was interpreted as arising from a CO column density of 1.0$^{+4.0}_{-0.5}\times 10^{20}$\,cm$^{-2}$, very consistent with our result (1.1$\pm$0.3$\times 10^{20}$\,cm$^{-2}$). Possibly, both studies are very affected by saturation of the lines. Combining the data into one excitation model is poorly justified by the too long time span between the observing dates that could affect the temperatures and sizes of the emission regions. Nevertheless, with our CASSIS excitation model we calculated the spectrum near 163\,$\mu$m and compared it directly to the PACS spectrum covering the CO 16--15 and H$_2$S 6,0,6--5,1,5 lines presented in \citet{exter}. Within observational uncertainties, the model reproduces the spectrum satisfactorily well, suggesting the temperatures and column densities of the gas did not change much in V838\,Mon's remnant within the last few years.

\end{appendix}

\end{document}